\documentclass[showpacs]{revtex4}

\usepackage{amsmath,latexsym,graphicx,subfigure,amssymb,amsfonts,mathbbol,%
float,times, mathptm,bm}

\renewcommand{\v}[1]{\mathbf{#1}}
\newcommand{\C}[1]{\mathcal{#1}}
\newcommand{\PP}[0]{\mathcal{P}}
\newcommand{\co}[1]{\C{C}_{#1}}
\newcommand{\msf}[1]{\mathsf{#1}}

\newcommand{\bl}[0]{\bullet}
\newcommand{\ad}[0]{^{\dag}}
\newcommand{\bra}[0]{\langle}
\newcommand{\ket}[0]{\rangle}
\newcommand{\Bra}[0]{\left\langle}
\newcommand{\Ket}[0]{\right\rangle}
\newcommand{\lst}[0]{\left|}
\newcommand{\rst}[0]{\right|}
\newcommand{\wt}[1]{\widetilde{#1}}
\newcommand{\wh}[1]{\widehat{#1}}

\newcommand{\ZZ}[0]{\mathbb{Z}}

\renewcommand{\Re}[0]{\mathfrak{Re}\!~}
\renewcommand{\Im}[0]{\mathfrak{Im}\!~}

\renewcommand{\min}[0]{{\text{min}}}

\newlength{\minus}
\newcommand{\ms}[0]{\hspace{\minus}}

\begin{document}

\title{Collective states in highly symmetric atomic configurations,
and single-photon traps}
\author{Hanno Hammer}
\email{H.Hammer@umist.ac.uk}
\affiliation{Department of Mathematics \\ University of Manchester
Institute of Science and Technology (UMIST) \\ P.O. Box 88 \\
Manchester M60 1QD \\ United Kingdom}
\date{\today}

\begin{abstract}

We study correlated states in a circular and linear-chain
configuration of identical two-level atoms containing the energy of a
single quasi-resonant photon in the form of a collective excitation,
where the collective behaviour is mediated by exchange of transverse
photons between the atoms. For a circular atomic configuration
containing $N$ atoms, the collective energy eigenstates can be
determined by group-theoretical means, making use of the fact that the
configuration possesses a cyclic symmetry group $\mathbb{Z}_N$. For
these circular configurations the carrier spaces of the various
irreducible representations of the symmetry group are at most
two-dimensional, so that the effective Hamiltonian on the
radiationless subspace of the system can be diagonalized
analytically. As a consequence, the radiationless energy eigenstates
carry a $\mathbb{Z}_N$ quantum number $p=0,1, \ldots, N$ which is
analogous to the angular momentum quantum number $l= 0, 1, \ldots$,
carried by particles propagating in a central potential, such as a
hydrogen-like system. Just as the hydrogen $s$-states are the only
electronic wave functions which can occupy the central region of the
Coulomb potential, the quasi-particle corresponding to a collective
excitation of the circular atomic sample can occupy the central atom
only for vanishing $\mathbb{Z}_N$ quantum number $p$. When a central
atom is present, the $p=0$ state splits into two, showing
level-crossing at certain radii; in the regions between these radii,
damped oscillations between two "extreme" $p=0$ states occur, where
the excitation occupies either the outer atoms or the central atom
only. For large numbers of atoms in a maximally subradiant state, a
critical interatomic distance of $\lambda/2$ emerges both in the
linear-chain and the circular configuration of atoms. The spontaneous
decay rate of the linear configuration exhibits a jump-like "critical"
behaviour for next-neighbour distances close to a
half-wavelength. Furthermore, both the linear-chain and the circular
configuration exhibit exponential photon trapping once the
next-neighbour distance becomes less than a half-wavelength, with the
suppression of spontaneous decay being particularly pronounced in the
circular system. In this way, circular configurations containing
sufficiently many atoms may be natural candidates for {\it
single-photon traps}.

\vspace{1em}

\end{abstract}

\maketitle

\section{Introduction}

In the classical paper by Dicke \cite{Dicke1954a}, super- and
subradiance in a collection of two-level atoms was studied in the
near-field regime from a theoretical point of view. Within the same
near-field limit, the topic has subsequently been reviewed by a couple
of authors \cite{GrossHaroche1982a,BenedictEA1996a}, with a
comprehensive study of subradiance being given in the series of papers
\cite{CrubellierEA1985a,CrubellierEA1986a,CrubellierEA1987a,
CrubellierEA1987b}. While Wigner functions, squeezing properties and
decoherence of collective states in the near-field regime have been
presented in \cite{BenedictEA1996a,BenedictCzirjak1999a,
FoeldiEA2002a}, triggering of sub- and superradiant states was shown
to be possible in \cite{KeitelEA1992a}. An experimental observation of
super- and subradiance was reported in
\cite{PavoliniEA1985,DeVoeBrewer1996}. The near-field limit utilized
in these examples is called the "small-sample approximation", the term
deriving from the fact that the atoms are assumed to be so close
together that they are all subject to the same phase of the radiation
field. In this approximation, the interaction Hamiltonian is
independent of the spatial location of the atomic constituents,
similar to a long-wavelength approximation.

In contrast, our work presented here focuses on super-and subradiance
in highly symmetric atomic systems with arbitrarily large interatomic
distances. While in the first part of this report we deal with the
theory of simply-excited correlated states in an arbitrary sample of
atoms, in the second part we shall mainly be concerned with
subradiance and, in particular, with the radiation-trapping
capability, of circular and linear-chain atomic configurations. This
radiation trapping is a consequence of collective states of the atomic
sample which can have extremely small spontaneous decay rates as
compared to the single-atomic decay rate; these states therefore are
natural candidates for {\it single-photon traps}.

As mentioned above we are mainly interested in the decay of {\it
simply-excited} states of the atomic sample; these states may be
conceptualized as follows: Consider $N$ identical two-level atoms with
infinite mutual distances such that precisely one of the atoms is
excited, all others being in the ground state. If we now think of
adiabatically decreasing the distances, the atoms will begin to
interact with each other via their coupling to the radiation field,
and hence the excitation, which previously was localized at one atom
only, will distribute over the whole sample. The resulting states
therefore will be superpositions of the excited levels in the atoms in
such a way that the atomic sample still contains the energy of one
(more or less resonant) photon, but this energy is now delocalized
over the whole sample. Moreover, as a consequence of this
delocalization, the atoms will be strongly entangled with each other
as well as with the radiation field. Some of these correlated states
turn out to be capable of very strong suppression of spontaneous
decay; this radiation trapping is one of the main topics of this work.

We shall start with justifying our choice of gauge and the quantum
picture utilized in the quantization of the system atoms+radiation. We
then motivate a split of the state space into $0$- and $1$-photon
states which will be convenient for formulating the decay of a
correlated state in the collectively excited atomic sample. We shall
see how the $\ZZ_N$ symmetry of the configuration can be utilized to
diagonalize the effective Hamiltonian of the atomic degrees of freedom
on the subspaces carrying the irreducible representations of the
symmetry group $\ZZ_N$; since, for simply-excited states, these
subspaces are only one- or two-dimensional, the problem of
diagonalization then becomes trivial and can be carried out
analytically, producing both the complex eigenvalues, and the
associated eigenvectors, of the effective channel Hamiltonian. We then
use this theory to illustrate the mechanism behind super- and
subradiance; in particular we show that the divergence of level shifts
for vanishing interatomic distance is due to the Coulomb dipole-dipole
interaction only. It will be demonstrated that, for an ensemble with
$N$ outer atoms in a circular configuration, there are $(N-1)$ states
which are insensitive to the presence of a central atom: The
properties of the quasi-particle describing the collective excitation
do not change when the atom at the center is removed, simply because
the latter is not occupied. On the other hand, the two states in a
circular ensemble which do occupy the central atom are analogous to
the $s$-state wave functions in hydrogen-like systems: Just as the $s$
states carry angular momentum quantum number $l=0$ and therefore
transform under the identity representation of $SO(3)$, our collective
$p=0$ states correspond to the identity representation of the symmetry
group $\ZZ_N$; and just as the $s$ states are the only ones which are
nonvanishing at the center-of-symmetry of the Coulomb potential in a
hydrogen-like system, so are our $p=0$ states the only quasi-particle
states which occupy the central atom at the center-of-symmetry of the
circle. Quantum beats between two extreme $p=0$ states are possible,
one, in which only outer atoms are occupied, the other, in which only
the central atom is excited. At certain radii of the circle, the two
$p=0$ levels cross, and the beat frequency vanishes, making the
population transfer between the extreme configurations
aperiodic. Moreover, we show numerically that, by increasing the
number of atoms in the circle for fixed radius, we arrive at a domain
where the minimal spontaneous decay rate in the sample decreases
exponentially with the number of atoms in the circle. The associated
subradiant states of the atomic sample will therefore qualify as {\it
single-photon traps}. In this examination, a critical inter-atomic
distance of one-half of the dominant wavelength emerges. Finally, we
show that the same critical distance emerges in a linear-chain
configuration of atoms: here, it signifies a jump-like, almost
discontinuous, behaviour of the minimal spontaneous decay rate, from
close to zero to finite values.

\section{Hamiltonian of the system and electric-dipole picture}

We first formulate the decay of a simply-excited correlated state in a
sample of $N$ neutral identical two-level atoms, arranged in an
arbitrary planar pattern, such that all atomic dipole moments are
aligned perpendicularly to the plane. The atoms are labelled by
$A=1,2, \ldots, N$, each atom being localized around a center-of-mass
$\v{R}_A$ in space. These locations are regarded as fixed in the sense
that the associated center-of-mass degrees of freedom do not take part
in the dynamics; as a consequence, the quantities $\v{R}_A$ are
$c$-numbers, not operators. The atoms are assumed to be identical,
having a spatial extent on the order of magnitude of a Bohr radius
$a_0$, and consist of point charges which are labelled as
$q_{A\alpha}$. With respect to wavelengths associated with optical
transitions it is legitimate to ignore the spatial variation of the
electromagnetic field over the extension $a_0$ of each atom, so that
we may replace the field degrees of freedom
$\v{A}_{\bot}(\v{x}_{A\alpha},t)$ by $\v{A}_{\bot}(\v{R}_{\alpha},t)$.
This step constitutes the {\it long-wavelength approximation}. For
reasons of consistency we then must also replace the {\it
inter}-atomic Coulomb energy by its lowest-order multipole
approximation, which is the dipole-dipole energy $\sum_{A<B}
V^{\text{dip}}_{AB}$. After canonical quantization, the {\it
G\"oppert-Mayer transformation} yields the {\it electric-dipole
Hamiltonian}
\begin{subequations}
\label{su2}
\begin{gather}
 H = \sum_A \sum_{\alpha} \frac{\v{p}_{A\alpha}^2}{ 2 \, m_{A\alpha}}
 + \sum_A \sum_{\alpha < \beta} V_{A\alpha\beta} + \int d^3\!k\;
 \sum_{s=1}^2 \hbar \omega(k)\, a_s(\v{k}){\dag}\, a_s(\v{k}) -
 \label{su2a} \\
 - \sum_A \v{d}_A \bl \v{E}_{\bot}(\v{R}_A) = H_0 + H_I\label{su2b}
   \quad.
\end{gather}
\end{subequations}
Here the unperturbed Hamiltonian $H_0$, given in (\ref{su2a}),
contains the sum over atomic Hamiltonians $H_{0A}$ including the
Coulomb interaction $V_{A\alpha\beta}$ between the internal
constituents of atoms $A= 1, \ldots, N$, but without the Coulomb
dipole-dipole interaction $V^{\text{dip}}_{AB}$ between atoms $A$ and
$B$, for $A \neq B$; and the normally-ordered free field energy. Since
the charge ensembles $A=1,2,\ldots$ are assumed to have mutual
distances $R_{AB} = |\v{R}_A - \v{R}_B |$ which are much larger than
the typical extension $a_0$ of the atomic or molecular wavefunctions,
wavefunctions belonging to different ensembles do not overlap. As a
consequence, atomic operators associated with different ensembles
commute,
\begin{equation}
\label{su1a}
 [x_{A\alpha i}, p_{B\beta j}] = 0 \quad, \quad \text{for $A\neq B$}
 \quad.
\end{equation}
The electric-dipole interaction is given in (\ref{su2b}), where
$\v{E}_{\bot}$ denotes the transverse electric field operator. The
interatomic Coulomb dipole-dipole interaction $V^{\text{dip}}_{AB}$
seems to be conspicuously absent in (\ref{su2}); however, it is a
feature of the G\"oppert-Mayer transformation (and more generally, of
the Power-Zienau-Woolley transformation yielding leading to the
multipolar Hamiltonian) to transform this interaction into a part of
the transverse electric field, so that the Coulomb interaction emerges
in {\it fully retarded} form as a part of the level-shift operator on
the radiationless subspace of the system. This will be seen in
formulae~(\ref{proc26} -- \ref{proc28}) below, thus clarifying that
the {\it inter}-atomic Coulomb interaction $V^{\text{dip}}_{AB}$ is
certainly contained in the Hamiltonian~(\ref{su2}), albeit in a
nonobvious way.

\section{The decay of a collective atomic excitation}

The simply-excited {\it uncorrelated} states of the sample are the
product states $|e_1,g_2,\cdots,g_N,0\rangle \equiv |1,0\rangle$,
$\ldots$, $|g_1,g_2,\cdots,e_N,0\rangle \equiv |N,0\rangle$, where
$|g_A\rangle$, $|e_A\rangle$ denote the ground and excited level of
the $A$th atom, and $|0\rangle$ is the radiative vacuum. The states
$\lst e \Ket$ are supposed to be electronic excitations, corresponding
to the fact that the exchanged photons will have optical
wavelengths. In the discussion before eq.~(\ref{su1a}) we pointed out
that the typical interatomic distance in the sample should be
comparable to an optical wavelength, so that electronic wavefunctions
belonging to excitations of different atoms will certainly not
overlap; as a consequence we can refrain from using antisymmetrized
electronic states (Slater determinants) and use simple product states
to describe the atomic sample.

The simply-excited uncorrelated states $\lst A,0\Ket$ are coupled to
the continuum of one-photon states $|g_1,g_2,\cdots,g_N,{\bf
k}s\rangle \equiv |G, \v{k}s \rangle$, where $({\bf k}s)$ is the wave
vector and $s$ denotes the polarization of the photon. As mentioned
above, the simply-excited correlated states will be superpositions of
the form
\begin{equation}
\label{coll1}
 \lst \C{C}\Ket = \sum_{B=1}^N c_B\, |B,0\rangle \quad,
\end{equation}
where the complex coefficients $c_A$ are to be determined from the
condition that the states~(\ref{coll1}) be energy eigenstates of a
suitable effective Hamiltonian. These states are formally very similar
to Frenkel excitons \cite{Frenkel1931a,Davydov,Knox,KenkreReineker},
e.g. in molecular crystals.

We are interested in the spontaneous radiative decay of a
simply-excited correlated atomic state which is coupled to a continuum
of one-photon states. The dominant contribution to this decay will
come from a quasi-resonant single-photon transition so that
two-photon- or higher-photon-number processes can be expected to play
a negligible role, and hence it will be admissible to truncate the
possible quantum states of the radiation field to one-photon states
$\lst \v{k}s\Ket$, and the vacuum $|0\rangle$. The
electric-dipole-admissible single-photon transitions of our correlated
atomic sample are
\begin{subequations}
\label{proc1}
\begin{align}
 \lst \C{C}\Ket & \rightarrow \lst G, \v{k}s\Ket \quad, \label{proc1a}
 \\
 \lst \C{C}\Ket & \rightarrow \lst AB, \v{k}s\Ket
 \quad. \label{proc1b}
\end{align}
\end{subequations}
where, in the second line~(\ref{proc1b}), the sample emits a photon
and makes a transition to a doubly-excited state $\lst AB\Ket$. Such
processes are inherently non-resonant and will be neglected, a step
which is usually called the {\it rotating-wave approximation}
\cite{Loudon}. We therefore take into account only transitions of the
sample to the common ground state, accompanied by emission of a single
photon (\ref{proc1a}). Thus, the state space of our joint system
atoms+radiation is spanned by the simply-excited radiationless states
$\lst A,0\Ket$, and the continuum of one-photon states $\lst G,
\v{k}s\Ket$, both of which are eigenstates of the unperturbed
Hamiltonian $H_0$. We now split the state space into a radiationless
subspace, spanned by the simply-excited states $\lst A,0\Ket$, and the
subspace of one-photon states $\lst G, \v{k} s\Ket$; the associated
projectors are
\begin{equation}
\label{proQ1}
 Q = \sum_{A=1}^N \lst A,0\Ket \Bra A,0\rst
\end{equation}
and $P = \int d^3\!k\; \sum_s \lst G, \v{k} s\Ket \Bra G, \v{k} s
\rst$. By construction, $Q$ and $P$ commute with $H_0$.

We now formulate the decay  of a given simply-excited correlated state
$\lst \C{C}\Ket$ into the continuum of one-photon states: Let $U(t,0)$
be the evolution operator  associated with the Hamiltonian~(\ref{su2})
in the Schr\"odinger  picture, and let us assume  that, at time $t=0$,
the atomic sample is in a correlated state, the radiation field is in
the vacuum  state $|0\rangle$, and  no correlations between  atoms and
radiation are present, so that the state vector of the total system is
$\lst  \psi_{t=0}\Ket  = \lst  \C{C}\Ket$. We wish to compute the
probability that the radiation at time $t>0$ is still trapped in the
system; in other words, the probability that the system at time $t$
can be found within the radiationless $Q$-space,
\begin{equation}
\label{amp2}
 P(t) = \sum_{A=1}^N \Bra \C{C}\rst U\ad(t,0) \lst A,0\Ket \Bra
 A,0\rst U(t,0) \lst \C{C}\Ket \quad.
\end{equation}
The evolution operator $U(t,0)$ can be expressed as a Fourier
transform over the Green operator, so that, for $t>0$,
\begin{equation}
\label{amp3}
 \Bra A,0\rst U(t,0) \lst \C{C}\Ket = - \sum_{B=1}^N \frac{1}{2\pi i}
 \int dE\; e^{-\frac{i}{\hbar} Et}\; \cdot \Bra A,0\rst\, QG(E_+)Q\,
 \lst B,0\Ket\; c_B \quad,
\end{equation}
where we have used eq.~(\ref{coll1}). The $Q$-space Green operator $Q
G(E_+) Q$ can be computed by standard methods as
\begin{equation}
\label{amp4}
 QG(z)Q = Q\bigg( z-H_0-Q H_I Q - Q H_I P \left(z- H_0 - P H_I P
 \right)^{-1} P H_I Q \bigg)^{-1} \quad.
\end{equation}
In the present case, the $P$-space is spanned by one-photon states
only, so that $P H_I P =0$; and similarly, $Q H_I Q = 0$. The
energy-dependent non-Hermitean Hamiltonian in the $Q$-channel
\begin{equation}
\label{chde10}
 Q \C{H}(z) Q = H_0 + Q H_I P \left(z-H_0\right)^{-1} P H_I Q \quad
\end{equation}
has left and right eigenvectors
\begin{subequations}
\label{chde11}
\begin{align}
 \Bra p^*, z\rst\, \C{H}(z) & = \Lambda_p(z)\, \Bra p^*,z\rst \quad,
 \label{chde11a} \\
 \C{H}(z)\, \lst p ,z\Ket & = \Lambda_p(z)\, \lst p,z\Ket \quad
 \label{chde11b}
\end{align}
which satisfy a generalized orthonormality relation
\cite{MorseFeshbach1+2}
\begin{equation}
\label{chde11c}
 \Bra p^*,z \rst \left. q, z\Ket = \delta_{pq} \quad,
\end{equation}
and which are $Q$-space-complete in the sense that
\begin{equation}
\label{chde11d}
 Q = \sum_p \lst p,z\Ket \Bra p^*, z\rst \quad.
\end{equation}
\end{subequations}
The projector property, $Q^2=Q$, immediately follows from
(\ref{chde11c}) and (\ref{chde11d}). 

With the help of relations (\ref{chde11}) we can write the
non-Hermitean channel Hamiltonian in the form
\begin{equation}
\label{chde12A}
 \C{H}(z) = \sum_p \lst p,z\Ket\, \Lambda_p(z)\, \Bra p^*,z\rst \quad,
\end{equation}
which is obviously a generalization of the diagonal form of Hermitean
Hamiltonians in the associated basis of energy eigenvectors. The
$Q$-space Green operator can be expressed similarly,
\begin{equation}
\label{chde12}
 Q G(z) Q = \bigg( z- \C{H}(z) \bigg)^{-1} = \sum_{p=1}^N \frac{\lst
 p,z\Ket \Bra p^*,z\rst}{z- \Lambda_p(z)} \quad.
\end{equation}
If this is inserted into (\ref{amp3}) for $z=E+i\epsilon$ we obtain an
expression for the transition amplitude which is still exact. However,
in the neighbourhood of the unperturbed initial energy $E_i$,
\begin{subequations}
\label{unpin1}
\begin{align}
 E_i & = (N-1) \cdot E_g + E_e = E_G + E_{eg} \quad,
\label{unpin1a} \\
 E_G & = N \cdot E_g \quad, \label{unpin1b} \\
 E_{eg} & = E_e - E_g \equiv \hbar c k_{eg} \quad. \label{unpin1c}
\end{align}
\end{subequations}
the variation of $Q G(z) Q$ near $z=E+i\epsilon$ due to the
quasi-resonant behaviour of the denominator can be expected to be much
stronger than the variation due to the functional dependence $z
\mapsto \C{H}(z)$; for this reason, $\C{H}(z)$ can be replaced by its
value $\C{H}(E_{i+})$ at the dominant energy of the process under
consideration. Then (\ref{chde12}) becomes
\begin{equation}
\label{chde13}
 Q G(z) Q \simeq \bigg( z- \C{H}(E_{i+}) \bigg)^{-1} = \sum_{p=1}^N
 \frac{\lst p,E_{i+}\Ket \Bra p^*,E_{i+}\rst}{z- \Lambda_p(E_{i+})}
 \quad.
\end{equation}
We need the matrix representation of this operator in the uncorrelated
basis $\lst A,0\Ket$. Let us abbreviate
\begin{equation}
\label{proc32}
 \lst \co{p} \Ket \equiv \lst p, E_{i+}\Ket \quad, \quad \Bra \co{p}^*
 \rst \equiv \Bra p^*, E_{i+}\rst \quad, \quad \Lambda_p \equiv
 \Lambda_p(E_{i+}) \quad,
\end{equation}
then the matrix of (\ref{chde13}) in the basis $\lst A,0\Ket$ is given
by
\begin{gather}
 \Bra A,0\rst Q G(E_{+}) Q \lst B,0\Ket = \sum_{p=1}^N \frac{\Bra
 A,0\right. \lst \co{p} \Ket \Bra \co{p}^* \rst \left. B,0\Ket}{E_+
 -\Lambda_p} \quad. \label{proc33}
\end{gather}
When (\ref{proc33}) is inserted into (\ref{amp3}), the Fourier
integral can be performed by means of the method of residues, since
the eigenvalues $\Lambda_p$ are now energy-independent. The
eigenvalues are generally complex, all imaginary parts being {\it
negative},
\begin{equation}
\label{proc35}
 \Im \Lambda_p < 0 \quad \text{for all $p=1,\ldots, N$} \quad,
\end{equation}
since all states in the $Q$-space have a finite lifetime and therefore
must eventually decay; this carries over into the
condition~(\ref{proc35}). The result for the transition amplitude is
then
\begin{gather}
 \Bra A,0\rst U(t,0) \lst \C{C}\Ket = \sum_{Bp} \Bra A,0\right. \lst
 \co{p} \Ket \Bra \co{p}^* \rst \left. B,0 \Ket\; c_B\;
 e^{-\frac{i}{\hbar} \Lambda_p t} \quad. \label{amp5AA1}
\end{gather}

\begin{figure}[htb]
\begin{minipage}[b]{0.55\textwidth}
   \centering
   \includegraphics[width=1.\textwidth]{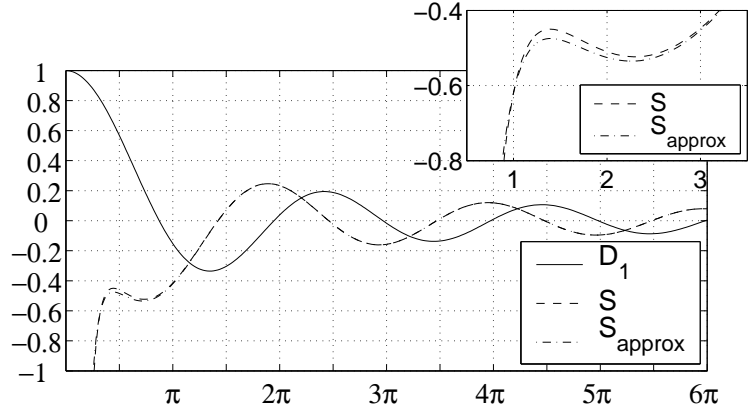}
\end{minipage}
\caption{Plots of the correlation functions $D_1(X)$ and $S(X)$ which
are responsible for spontaneous decay (''D'') and level shifts
(''S''), respectively. $D_1$ becomes maximal at zero interatomic
distance while $S$ tends to minus infinity there, due to Coulomb- and
radiative self-energy of the atomic dipole. Both functions tend to
zero as $X \rightarrow\infty$, expressing the fact that any kind of
correlation must cease to exist for infinite distance. The inset shows
a comparison between the exact shift function $S(X)$, based on
eq.~(\ref{eig1b}), and the approximation $S_{\text{approx}}$, based on
eq.~(\ref{appr2}), respectively. \label{figure2}}
\end{figure}

We now go back to equation (\ref{chde10}) for the non-Hermitean
$Q$-channel Hamiltonian. At the dominant energy $E_{i+} = E_i +
i\epsilon$ we can write
\begin{gather}
 \C{H}(E_{i+}) = Q H_0 Q + \hbar \Delta(E_i) - i\, \frac{\hbar}{2}\,
 \Gamma(E_i) \quad,
 \label{proc2}
\end{gather}
where $\Delta(E_i)$ contains the level shifts in the $Q$-space, while
$\Gamma(E_i)$ contains the (spontaneous) decay rates. In terms of the
uncorrelated basis $\lst A,0\Ket$ we have
\begin{subequations}
\label{proc3}
\begin{equation}
\label{proc3C}
 \Bra A,0 \rst \C{H}(E_+) \lst B,0\Ket  = E_i\, \delta_{AB} + \hbar
 \Delta_{AB}(E_i) - i\, \frac{\hbar}{2}\, \Gamma_{AB}(E_i) \quad,
\end{equation}
where
\begin{align}
 \hbar \Delta_{AB}(E) & \equiv \PP \int d^3\!k \sum_{s=1}^2 \frac{
 \Bra A,0 \rst H_I \lst G, \v{k} s\Ket \Bra G, \v{k} s\rst H_I \lst B,
 0\Ket}{E - E_G - \hbar \omega} \quad, \label{proc3a} \\
 \frac{\hbar}{2}\, \Gamma_{AB}(E) & \equiv \pi \int d^3\!k \sum_{s=1}^2\; 
 \Bra A,0 \rst H_I \lst G, \v{k} s\Ket \Bra G, \v{k} s\rst H_I \lst B,
 0\Ket\; \delta(E - E_G - \hbar \omega) \quad. \label{proc3b}
\end{align}
\end{subequations}
The initial energy is $E_i = E_G + \hbar c k_{eg}$, where $k_{eg}$ is
defined in (\ref{unpin1c}). The decay matrix at the initial
energy can be computed by integrating out the delta function,
\begin{subequations}
\label{proc13}
\begin{align}
 \Gamma_{AB}(E_i) & = \Gamma\, D_1( k_{eg} R_{AB}) \quad,
 \label{proc13a} \\
 \Gamma & \equiv \frac{d^2\, k_{eg}^3}{3\pi \epsilon_0 \hbar} \quad,
 \label{proc13b} \\
 D_1(X) & = \frac{3}{2} \left( \frac{\sin X}{X} + \frac{\cos X}{X^2} -
 \frac{\sin X}{X^3} \right) \quad, \label{proc13c}
\end{align}
\end{subequations}
where $\Gamma$ is the spontaneous emission rate of a {\it single} atom
in a radiative vacuum. The argument of the function $D_1$ is equal to
$2\pi$ times the distance $R_{AB}$ between atoms $A$ and $B$ measured
in units of the wavelength $\lambda_{eg} = 2\pi/k_{eg}$ of the Bohr
transition $|e\rangle \rightarrow |g\rangle$. A plot of the function
$X \mapsto D_1(X)$ is given in Fig.~\ref{figure2}.

[Remark: In the more general case of arbitrarily aligned atoms there
are two correlation functions $D_1, D_2$ emerging in the decay matrix
rather than just one. This is indicated by our denotation. For
parallely aligned atoms, the contribution of $D_2$ vanishes.]

The level-shift matrix $\Delta_{AB}$ and the decay matrix
$\Gamma_{AB}$ are connected by a dispersion relation which, for the
case at hand, reads
\begin{equation}
\label{proc5}
 \Delta_{AB}(E) = \frac{1}{2\pi} \int\limits_0^{\infty} dE'\; \PP
 \frac{\Gamma_{AB}(E_G+E')}{E -E_G -E'} \quad.
\end{equation}
The energy $E_G = N\cdot E_g$ of the unperturbed joint atomic ground
state will be negative in general. Using (\ref{proc5}) the level
shifts can be computed using appropriate complex contour integrals,
with the result
\begin{subequations}
\label{eig1}
\begin{align}
 \Delta_{AB}(E_i) & = -\frac{1}{2} \Gamma \, S(X) \quad, \quad X =
 k_{eg} R_{AB} \label{eig1a} \\
 S(X) & = \frac{3}{2} \left\{ \frac{\cos X}{X} - \frac{ \sin X}{X^2} -
 \frac{ \cos X}{X^3} + \frac{1}{\pi X^2} \int\limits_0^{\infty} du\;
 \frac{ e^{-u} \left( 1 + u + u^2 \right)}{ u^2 + X^2} \right\}
 \quad. \label{eig1b}
\end{align}
\end{subequations}
The integral on the right-hand side of~(\ref{eig1b}) can be
approximated as follows: We replace the exponential $e^{-u}$ by $1$
for $0 \le u \le 1$; and by $0$ for $u> 1$. This leads to the
following approximation for the function $S(X)$:
\begin{equation}
\label{appr2}
  S(X)_{\text{approx}} \simeq \frac{3}{2} \left\{ \frac{\cos X}{X} -
 \frac{ \sin X}{X^2} - \frac{ \cos X}{X^3} + \frac{1}{\pi X^2} \left[
 \frac{1- X^2}{X}\, \arctan \frac{1}{X} + 1 + \frac{1}{2}\, \ln
 \frac{1 +X^2}{X^2} \right] \right\} \quad.
\end{equation}
A plot of the exact function $S(X)$ as given in~(\ref{eig1b}), and its
approximation~(\ref{appr2}), is given in Fig.~\ref{figure2}.

The spatial dependence of the correlations between the atoms is now
fully contained in the functions $D_1(X)$ (''decay'') and $S(X)$
(''shift''). Both functions tend to zero as $X \rightarrow\infty$;
this is clear on physical grounds, since $X= \frac{2\pi|{\bf R}_A
-{\bf R}_B|}{\lambda_{eg}}$ implies that the interatomic distance
tends to infinity in this limit, in which case all correlations
between the atoms must cease to exist. The only exception of course is
the self-correlation of atom $A$, as expressed by the fact that the
diagonal elements $\Gamma_{AA}$ are equal to one at each
distance. Thus, the decay matrix $\Gamma_{AB}$ tends to $\Gamma \cdot
\Eins$ as $X \rightarrow \infty$, implying that the spontaneous
emission rate for any simply-excited collective state $\lst \C{C}\Ket$
tends to the single-atom rate $\Gamma$, as expected.

Now we study the opposite limit: The decay matrix $\Gamma_{AB}$ has a
well-defined limit for vanishing interatomic distance, since
\begin{equation}
\label{eig1B1}
 \lim\limits_{X\rightarrow 0} D_1(X) = 1 \quad,
\end{equation}
and as a consequence, $\Gamma_{AB} \rightarrow \Gamma \;\; \forall
A,B$ in this limit. The reason for this behaviour is as follows: The
limit $X \rightarrow 0$ defines the {\it small-sample limit}, in which
all atomic dipoles are within each other's near zone. It is easy to
compute that, in this limit, their degree of correlation is
proportional to the cosine ${\bf e}_A \bullet{\bf e}_B$, where
$\v{e}_A$ and $\v{e}_B$ are unit vectors in the direction of the
atomic dipoles. Thus, in the small-sample approximation, parallely
aligned dipoles are maximally correlated while dipoles oriented
perpendicularly are not correlated at all. In our sample, all dipoles
are parallel, hence the limit (\ref{eig1B1}).

On the other hand, the function $S(X)$ diverges in the limit $X
\rightarrow 0$. The reason for this divergence is different depending
on whether off-diagonal or diagonal elements of $\Delta_{AB}$ are
considered. Let us first study the off-diagonal case:

Here, $R_{AB} \neq 0$, and we can rewrite (\ref{eig1a}) in the form
\begin{gather}
 \Delta_{AB}(E_i)  = \frac{d^2}{4\pi\epsilon_0 \hbar}
 \frac{1}{R_{AB}^3} \; f(R_{AB}) \quad, \label{proc26}
\end{gather}
where $f(R_{AB})$ is an analytic function of $R_{AB}$ which tends to
$\frac{1}{2}$ as $R_{AB}$ tends to zero. This last result should be
compared with the Coulomb dipole-dipole interaction between the two
atoms $A$ and $B$,
\begin{subequations}
\begin{equation}
\label{proc27a}
 V^{\text{dip}}_{AB} = \frac{ \v{d}_A \bl \v{d}_B - 3\, \left( \v{d}_A
 \bl \v{e}_{AB} \right) \left( \v{d}_B \bl \v{e}_{AB} \right) }{ 4\pi
 \epsilon_0 \, R_{AB}^3 } \quad.
\end{equation}
For two classical dipoles which are parallel, $\v{d}_A \bl \v{d}_B =
d^2$, and which are aligned perpendicularly to their axis $\v{e}_{AB}$
of connection, $\v{d}_A \bl \v{e}_{AB} = \v{d}_B \bl \v{e}_{AB} = 0$,
(\ref{proc27a}) takes the simpler form
\begin{equation}
\label{proc27b}
 V^{\text{dip}}_{AB}(R_{AB}) = \frac{d^2 }{ 4\pi \epsilon_0\,
 R_{AB}^3} \quad.
\end{equation}
\end{subequations}
It then follows that the level shifts in (\ref{proc26}) contain the
dipole-dipole Coulomb interaction between the charged ensembles $A$
and $B$, since (\ref{proc27b}) is contained as a factor in the
off-diagonal matrix element $\hbar \Delta_{AB}(E_i)$ describing the
level shift of the system due to the coupling to the radiative degrees
of freedom {\it as well as} the dipole-dipole Coulomb interaction
between atoms $A$ and $B$. So we could indeed write
\begin{equation}
\label{proc28}
 \hbar \Delta_{AB}(E_i) = V^{\text{dip}}_{AB}(R_{AB}) \cdot f(R_{AB})
 \quad,
\end{equation}
where $f$ now contains the effect of the purely radiative degrees of
freedom. As mentioned above, in the limit $R_{AB} \rightarrow 0$ we
have $f(R_{AB}) \rightarrow \frac{1}{2}$; eq. (\ref{proc28}) then
implies that
\begin{equation}
\label{proc28A}
 \lim_{R_{AB} \rightarrow 0} \hbar \Delta_{AB}(E_i) = \frac{1}{2}
 \lim_{R_{AB} \rightarrow 0} V^{\text{dip}}_{AB}(R_{AB}) \quad, \quad
 A \neq B \quad.
\end{equation}

We now turn to discuss the behaviour of the diagonal elements
$\Delta_{AA}$: These quantities are infinite, since $R_{AA}=0$. This
divergence remains even if only a single atom is considered, and has
to do with the inevitable coupling of the atom to the longitudinal and
transverse degrees of freedom of the radiation field. From
(\ref{proc26}~--~\ref{proc28}) we learn that part of the divergence
comes from the Coulomb dipole self-energy of atom $A$. The other
contribution to the self-energy is associated with emission and
reabsorption of virtual photons of the dipole which is an ongoing
process even if only one atom is embedded in the vacuum. It emerges
explicitly as a radiative dipole self-energy (an infinite $c$-number
term), when the G\"oppert-Mayer transformation is performed on the
Standard Hamiltonian. These processes are usually summarized by saying
that the energy levels, together with the states, undergo {\it
radiative corrections}.

As a consequence, the unperturbed states are not really the true
physical states; rather, they are abstract constructions whose
coupling to the continuum of radiation modes shifts their unperturbed
energy by an infinite, but {\it unobservable} amount, an effect which
is well-known in Quantum Field Theory \cite{BjorkenDrell1,
BjorkenDrell2, BogoliubovShirkov, ItzyksonZuber, RyderQFT,
BailinLoveGFT}. This amount must be absorbed into the definition of
the unperturbed energy, so that the energy of the initial state $\lst
\psi_{t=0}\Ket = \lst \C{C} \Ket$ becomes {\it renormalized}
\begin{equation}
\label{eig1B1a}
\begin{aligned}
 E_i & \longrightarrow E_i + \lim\limits_{X \rightarrow 0} \hbar
 \Delta_{AA}(E_i) \equiv \widetilde{E}_i \quad,
\end{aligned}
\end{equation}
where $\widetilde{E}_i$ is now assumed to have a finite value, namely
the value unperturbed by the presence of $N-1$ other atoms; as a
consequence, $E_i$ must be assumed to have been infinite in the first
place. In turn, we now must subtract the same infinite quantity from
the diagonal elements of the level-shift matrix, which amounts to a
redefinition
\begin{equation}
\label{eig1B2}
 \Delta_{AB}(E_i) \longrightarrow \widetilde{\Delta}_{AB}(E_i)
 \equiv \begin{cases} \quad \Delta_{AB}(E_i) & , \quad A \neq B
 \; , \\ \quad 0 & , \quad A=B \; . \end{cases}
\end{equation}
As a consequence, the only observable level-shifts are now those due
to {\it inter}atomic interactions.

Using the  functions $D_1$ and $S$, the  matrix element (\ref{proc3C})
can be written as
\begin{gather}
 \Bra A,0\rst \C{H}(E_i) \lst B,0\Ket = \left\{
\begin{array}{c@{}cl@{}c}
 - \frac{\hbar \Gamma}{2}\, \bigg[ \, S(k_{eg} R_{AB}) + i\,
 D_1(k_{eg} R_{AB}) \, \bigg] & \; , & A \neq B  & \; , \\[10pt]
 \widetilde{E}_i - i\, \frac{\hbar \Gamma}{2} & \; , & A=B & \; .
\end{array}
\right\} = \widetilde{E}_i\, \delta_{AB} - \frac{\hbar \Gamma}{2}\,
\C{R}_{AB} \quad,
\label{eig2}
\end{gather}
where the matrix elements $\C{R}_{AB}$ are determined by the function
$M(X) = S(X) + i D_1(X)$, which can be determined from (\ref{proc13c})
and (\ref{appr2}) to be
\begin{equation}
\label{eig7}
 M(X) \equiv S(X)+i\, D_1(X) \simeq \frac{3}{2}\, e^{iX}\, \left(
 \frac{1}{X} + \frac{i}{X^2} - \frac{1}{X^3} \right) + 
\frac{3}{2 \pi X^2} \left( \frac{1- X^2}{X}\, \arctan \frac{1}{X} + 1
 + \frac{1}{2}\, \ln \frac{1 +X^2}{X^2} \right) \quad.
\end{equation}
Evidently, if we have found the left/right eigenvectors of $\C{R}$,
then we have diagonalized the whole channel Hamiltonian, for $\mu_p$
is an eigenvalue of $\C{R}$ if and only if $\wt{E}_i -\frac{\hbar
\Gamma}{2}\,\mu_p$ is an eigenvalue of $\widetilde{E}_i\, \Eins_N -
\frac{\hbar \Gamma}{2}\, \C{R}$. It follows from eq.~(\ref{chde12A})
that the matrix of the channel Hamiltonian $\C{H}(E_i)$ in the basis
$\lst A,0\Ket$ can be written as
\begin{equation}
\label{eig3A}
 \Bra A,0 \rst \C{H}(E_i) \lst B,0\Ket = \sum_{p=1}^N \Bra A,0\rst
 \left. \co{p} \Ket \bigg\{ \wt{E}_i -\frac{\hbar
 \Gamma(k_{eg})}{2}\,\mu_p \bigg\} \Bra \co{p}^* \rst \left. B,0\Ket
 \quad.
\end{equation}
We must be aware that $\wt{E}_i$ has been renormalized by an infinite
amount and is now to be regarded as the finite, physically meaningful
energy of the simply-excited atomic sample when all interatomic
distances are infinite. A comparison of (\ref{eig3A}) with
(\ref{proc3C}) now shows that the quantities $-\frac{\hbar \Gamma}{2}
\mu_p$ are the right eigenvalues of the matrix
\begin{equation}
\label{eig3B}
 \hbar \wt{\Delta}_{AB}(E_i) - i \frac{\hbar}{2} \Gamma_{AB}(E_i)
 \quad.
\end{equation}
It is therefore appropriate to denote the real and imaginary parts of
these eigenvalues according to
\begin{equation}
\label{eig5}
 -\frac{\hbar\Gamma}{2} \bigg( \Re \mu_p + i \Im \mu_p \bigg)
  = \hbar \wt{\Delta}_p - i \frac{\hbar}{2} \Gamma_p \quad,
\end{equation}
and hence
\begin{subequations}
\label{eig6}
\begin{align}
 \wt{\Delta}_p & = -\frac{\Gamma}{2}\; \Re\mu_p \quad,
 \label{eig6a} \\
 \Gamma_p & = \ms \Gamma\; \Im\mu_p \quad, \label{eig6b}
\end{align}
\end{subequations}
where $\wt{\Delta}_a$ now denotes the renormalized level-shifts. A
comparison of (\ref{eig3A}) with (\ref{chde12A}), on the other hand,
shows that the quantities in curly brackets in (\ref{eig3A}) are equal
to $\Lambda_p$. From (\ref{eig3A}, \ref{eig3B}, \ref{eig5}) it then
follows that $\Lambda_p$ should be decomposed according to
\begin{equation}
\label{amp6}
 \Lambda_p = \, \wt{E}_i + \hbar \wt{\Delta}_p - i\frac{\hbar}{2}
 \Gamma_p \quad, \quad \Gamma_p > 0 \quad,
\end{equation}
where all imaginary parts $\Gamma_p$ must be positive due to
(\ref{proc35}).

We now insert (\ref{amp6}) into equation (\ref{amp5AA1}) for the
transition amplitude and compute our target quantity, namely the
probability $P(t)$ for the atomic sample at time $t>0$ to be found in
the radiationless $Q$-space, as defined in (\ref{amp2}); the result is
\begin{subequations}
\label{amp5}
\begin{align}
 P(t) & = \sum_{B_1, B_2} c^*_{B_2}\; P_{B_2 B_1}\; c_{B_1} \quad,
 \label{amp5a} \\
 P_{B_2 B_1} & = \sum_{A,p_1,p_2=1}^N \Bra B_2, 0 \rst
 \left. \co{p_2}^* \Ket \Bra \co{p_2} \rst \left. A,0\Ket\; e^{i\left(
 \wt{\Delta}_{p_2} - \wt{\Delta}_{p_1} \right) t} \nonumber \\
 & \times e^{-\frac{1}{2} \left( \Gamma_{p_1} + \Gamma_{p_2} \right)
 t}\; \Bra A,0\rst \left. \co{p_1} \Ket \Bra \co{p_1}^* \rst
 \left. B_1,0\Ket \quad. \label{amp5b}
\end{align}
\end{subequations}

For the given radius $r$, number of atoms $N$, and configuration (a)
or (b), let $\Lambda_{\min}$ be the eigenvalue with the smallest decay
rate,
\begin{equation}
\label{amp10}
 \Gamma_{\min} \le \Gamma_p \quad \text{for all $p=1, \ldots, N$}
 \quad.
\end{equation}
Then the associated correlated state
\begin{equation}
\label{amp11}
 \lst \C{C} \Ket = \frac{1}{\sqrt{\Bra \co{\min} | \co{\min} \Ket}}\, \lst
 \co{\min} \Ket \quad
\end{equation}
has the longest lifetime with respect to spontaneous decay [saying
nothing about stability against environmental perturbations], and
hence is a candidate for a single-photon trap.

\section{Cyclic symmetry of the circular atomic configurations}

\begin{figure}[htb]
\begin{minipage}[b]{0.35\textwidth}
   \centering \includegraphics[width=1.\textwidth]{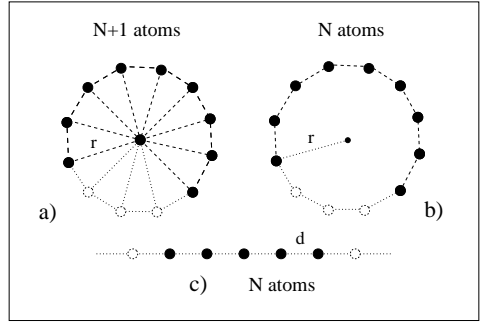}
\end{minipage}
\caption{The configurations of $N(+1)$ atoms which are examined: In
configuration~(a), $N$ outer atoms form a regular $N$-polygon with
radius $r$, plus an atom at the center. Configuration (b) is the same
as a), but without the central element. Configuration (c) has $N$
atoms in a linear chain with next-neighbour distance
$d$. \label{figure1}}
\end{figure}

In the remaining part of this paper we shall apply the above theory
mainly to circular configurations of atoms, containing $N$ atoms along
the perimeter of the circle, and an optional atom at the center. The
system with [without] central atom is referred to respectively as
configuration a) [ b)], see Fig.~\ref{figure1}. In the last section we
briefly give some results about configuration c), involving $N$ atoms
arranged along a straight line.

Now let us study the circular configurations: We shall construct the
eigenvectors $\lst \co{p} \Ket$ of $\C{H}(\wt{E}_i)$ explicitly by
group-theoretical means, taking advantage of the fact that the system
has a cyclic symmetry group
\begin{equation}
\label{cyclic1}
 G = \left\{ e, T, T^2, \ldots, T^{N-1} \right\} \quad, \quad T^{N} =
 e \quad,
\end{equation}
where the generator $T$ is realized by a (passive) rotation at the
angle $2\pi/N$ about the symmetry axis of the circle, and $N$ is the
number of outer atoms along the perimeter of the circle. This holds
for both configurations a) and b). All cyclic groups of the same order
$N$ are isomorphic to the group of integers $\ZZ_N = \{0, 1, \ldots, N
-1\}$ with group operation $A \circ B = A+B \mod N$, unit element $0$,
and inverses $A^{-1} = - A \mod N$. The set of all {\it unitary
irreducible} representations of $\ZZ_N$ is given by
\begin{equation}
\label{cyclic2}
 \Gamma^p\left( T^A \right) = \exp\left(\frac{2\pi i p A}{N} \right)
 \quad, \quad A = 1,2, \ldots, N \quad,
\end{equation}
where $ p = 0, 1, \ldots, N-1$. The atoms are located at the center
and along the perimeter of a circle with radius $r$ such that
\begin{equation}
\label{cyclic4}
\begin{aligned}
 \v{R}_z & = (0,0) \quad, \\
 \v{R}_A & = r\, \left( \cos\frac{2\pi (A-1)}{N},\, \sin\frac{2\pi
 (A-1)}{N} \right) \quad,
\end{aligned}
\end{equation}
for $A = 1, \ldots, N$.

As mentioned above, the basic symmetry operation in our system is the
shift $T$, corresponding to a {\it passive} rotation $T_c$ about the
symmetry axis by an angle $2\pi/N$. $T_c$ transforms the coordinates
$\v{R}_A$ of the outer atoms and the central atom $\v{R}_z$ according
to
\begin{equation}
\label{cyclic5}
\begin{aligned}
 T_c\, \v{R}_A & = \v{R}_{A-1} \quad, \quad N+1 = 1 \mod N \quad, \\
 T_c\, \v{R}_z & = \v{R}_z \quad.
\end{aligned}
\end{equation}
Here $\v{R}$ denotes the triple of coordinates with respect to an
orthonormal basis which undergoes rotation, and not the invariant
vector. Next we need to construct the action of the unitary operator
$T$ associated with the coordinate transformation $T_c$ given in
(\ref{cyclic5}) on the state vectors of the system. $T$ operates on
the atomic degrees of freedom only, so we can ignore the photon states
for the time being. It is sufficient to specify the action of $T$ on
product states; superscripts denote the label of the atoms:
\begin{equation}
\label{cyclic11}
 T\, \lst \psi^z_{i_z} \Ket \otimes \lst \psi^1_{i_1} \Ket \otimes
 \cdots \otimes \lst \psi^N_{i_N} \Ket = \lst \psi^z_{i_z} \Ket
 \otimes \lst \psi^1_{i_2} \Ket \otimes \cdots \otimes \lst
 \psi^{N-1}_{i_N} \Ket \otimes \lst \psi^N_{i_1} \Ket \quad.
\end{equation}
This action derives from eq.~(\ref{cyclic5}), and preserves the
subspaces of any given degree of excitation. On the single-excitation
subspace, $T$ acts according to
\begin{equation}
\label{cyclic12}
 T\, \lst A\Ket = \lst A-1 \Ket \quad, \quad T\, \lst z\Ket = \lst
 z\Ket \quad,
\end{equation}
and hence has the matrix elements
\begin{equation}
\label{cyclic13}
 T_{AB} = \Bra A \rst T \lst B\Ket = \delta_{A, B-1} \quad.
\end{equation}
The inverse is given by the transposed matrix, $T^{-1} = T^T$,
expressing the fact that $T$ is a unitary operator.

Eqs.~(\ref{cyclic12}) are consistent with the transformation behaviour
of a scalar wave function: Suppose that $\psi(\v{x}) = \Bra \v{x} |
\psi \Ket$ is a wave function in the coordinate system $\v{x}$, and
$T_c \v{x} = \v{x}'$ is a coordinate transformation. The wave function
is a scalar under the associated unitary transformation $T$ on the
state space if, in the new coordinates $\v{x}'$, the new state vector
$| \psi'\rangle$ with wave function $\psi'(\v{x}') = \langle T_c \v{x}
| T \psi \rangle$ satisfies
\begin{subequations}
\label{beisp1}
\begin{align}
 \psi'(\v{x}') & = \psi(\v{x}) \quad, \label{beisp1a}
\end{align}
and as a consequence
\begin{align}
 T\, \lst \v{x} \Ket & = \lst T_c\, \v{x} \Ket \quad. \label{beisp1c}
\end{align}
\end{subequations}
This should be compared with the case of the simply-excited correlated
states at hand: Here we can consider the sites $\v{R}_z, \v{R}_A$ of
the atoms as discrete locations analogous to the parameters $\v{x}$ in
(\ref{beisp1c}). The simply-excited state $\lst \C{C} \Ket$ may be
regarded as describing a scalar quasi-particle (a Frenkel exciton) for
which the states $|z\rangle, |A\rangle$ are the analogues of the state
vectors $|\v{x}\rangle$ for an ordinary spinless particle. The
transformations~(\ref{cyclic5}) then imply (\ref{cyclic12}), in full
analogy to (\ref{beisp1c}).

The channel Hamiltonians of both configurations are invariant under
this transformation,
\begin{equation}
\label{cyclic18}
 T^{-1}\, \C{H}\, T = \C{H} \quad, \quad T^{-1}\, \C{R}\, T = \C{R}
 \quad,
\end{equation}
where the matrix $\C{R}$ was defined in
eq.~(\ref{eig2}). Relations~(\ref{cyclic18}) can be proven as follows:
The matrix $\C{R}$ is symmetric by construction, and, for
configuration (a), has components
\begin{equation}
\label{cyclic21}
 \C{R} = \left( \begin{array}{c|cccc} i & \C{R}_{z1} & \C{R}_{z2} &
 \cdots & \C{R}_{zN} \\ \hline \C{R}_{1z} & i & \C{R}_{12} & \cdots &
 \C{R}_{1N} \\ \C{R}_{2z} & \C{R}_{21} & i & & \vdots \\ \vdots &
 \vdots & & \ddots & \vdots \\ \C{R}_{Nz} & \C{R}_{N1} & \C{R}_{N2} &
 \cdots & i \end{array} \right) \quad
\end{equation}
such that
\begin{subequations}
\label{cyclic22}
\begin{align}
 \C{R}_{z1} & = \C{R}_{z2} = \cdots = \C{R}_{zN} \quad,
 \label{cyclic22a} \\
 \C{R}_{A,B} & = \C{R}_{A+C, B+C} \quad \text{for} \quad 1 \le A, B
 \le N \quad, \label{cyclic22b}
\end{align}
\end{subequations}
where in (\ref{cyclic22b}) all indices are to be taken modulo $N$. The
relations (\ref{cyclic22}), in turn, follow from the fact that
\begin{subequations}
\label{cyclic23}
\begin{align}
 R_{z1} & = R_{z2} = \cdots = R_{zN} = r \quad, \label{cyclic23a} \\
 R_{AB} & = R_{A+C, B+C} \quad \text{for} \quad 1 \le A, B \le N
 \quad, \label{cyclic23b}
\end{align}
\end{subequations}
and arbitrary $C$, where $r$ is the radius of the circle. If the
matrix representation (\ref{cyclic13}) of the operator $T$ is used on
(\ref{cyclic21}) we obtain~(\ref{cyclic18}). Similarly, for the
configuration (b) without central atom, the relevant operators are
given by the lower right block matrices in eqs.~(\ref{cyclic13}) and
(\ref{cyclic21}), so that again (\ref{cyclic18}) holds. Thus, we have
proven that the generator $T$ of the cyclic group $\ZZ_N$ commutes
with the matrix $\C{R}$, and hence with the (non-Hermitean) channel
Hamiltonian $\C{H}(E_i)$, at least on the single-excitation subspace;
but these arguments can easily be extended to show that $T$ and
$\C{H}$ commute on all subspaces. Hence we have
\begin{equation}
\label{cyclic24}
 \Big[\, T, \, \C{R}\, \Big] = 0 \quad, \quad \Big[\, T, \, \C{H}\,
 \Big] = 0 \quad.
\end{equation}
Just as in the case of Hermitean operators, the commutativity
(\ref{cyclic24}) implies the existence of a common system of
eigenvectors of $\C{H}$ and $T$: Suppose that $v$ is an eigenvector of
$T$, regarded as a complex column vector of dimension $N$ or $(N+1)$,
with eigenvalue $t$, so that $T v = t v$. Then
\begin{equation}
\label{cyclic25}
 T\Big( \C{H} v \Big) = t \Big( \C{H} v \Big) \quad,
\end{equation}
where we have used (\ref{cyclic24}). Eq.~(\ref{cyclic25}) says that
$\C{H}$ preserves all eigenspaces of $T$, i.e. if $\msf{T}(t)$ is the
subspace corresponding to the eigenvalue $t$, then
\begin{equation}
\label{cyclic26}
 \C{H}\, \msf{T}(t) \subset \msf{T}(t) \quad.
\end{equation}
On the other hand we must have $T^A v = t^A v$, and in particular
\begin{equation}
\label{cyclic27}
 v = \Eins\, v = T^N\, v = t^N\, v \quad,
\end{equation}
from which it follows that
\begin{equation}
\label{cyclic28}
 t = \exp\left( \frac{2\pi i p}{N} \right) = \Gamma^p(T) \quad
\end{equation}
for some $p \in \{0, 1, \ldots, N-1\}$, and we see that the
eigenvalues of $T$ are just the one-dimensional matrices $\Gamma^p$,
eq.~(\ref{cyclic2}), of the irreducible representations of the cyclic
group $\ZZ_N$. Accordingly, we can label the eigenspaces of $T$ by the
index $p$ of the representation as $\msf{T}(t) \equiv \msf{T}_p$.
	
We see that the eigenvectors $v$ of $T$ in the state space of the
atomic system satisfy $T\, v = \Gamma^p(T)\, v$ and therefore span
carrier spaces for the irreducible representations (\ref{cyclic2}) of
the symmetry group. Since $T$ is unitary, these carrier spaces are
orthogonal, and their direct sum is the total atomic state
space. Since the channel Hamiltonian $\C{H}$ preserves the eigenspaces
$\msf{T}_p$ of $T$ according to (\ref{cyclic26}) we see that a method
to simplify the diagonalization of $\C{H}$ is given as follows:
\begin{enumerate}
\item
We first determine the eigenspaces $\msf{T}_p$ of $T$ on the state
space of the atomic degrees of freedom. These eigenspaces are
comprised by vectors $v$ each of which transforms as a basis vector
for a definite irreducible representation $\Gamma^p(T)$ of the
symmetry group $\ZZ_N$,
\begin{equation}
\label{cyclic29}
 T\, v = \Gamma^p(T)\, v \quad \text{for all $v \in \msf{T}_p$} \quad.
\end{equation}
\item
The channel Hamiltonian $\C{H}$ preserves each eigenspace $\msf{T}_p$,
so that we can diagonalize $\C{H}$ on each $\msf{T}_p$
separately.
\end{enumerate}
Since the eigenspaces $\msf{T}_p$ are smaller in dimension than the
original state space we can expect a significant simplification of the
diagonalization procedure for $\C{H}$. This will be explicitly
demonstrated in the following sections.

\section{Eigenspaces of the generator $T$ of the symmetry group}

We now construct the eigenspaces $\msf{T}_p$ of $T$ on the
single-excitation subspaces of the atomic systems with and without a
central atom. These eigenspaces may be constructed by a standard
procedure \cite{Cornwell1} by writing down the projection operators
$\C{P}^p$ onto $\msf{T}_p$ on the state space of atomic degrees of
freedom. These projectors have the following property: Given an
arbitrary element $|\C{C} \rangle$ of the (single-excitation) state
space, its image $\C{P}^p |\C{C}\rangle$ transforms like a basis
vector for the one-dimensional unitary irreducible representation
$\Gamma^p(T)$ defined in (\ref{cyclic2}),
\begin{equation}
\label{eigen1}
 T^A\, \Big( \C{P}^p \lst \C{C} \Ket \Big) = \Gamma^p\left( T^A
 \right)\, \Big( \C{P}^p \lst \C{C} \Ket \Big) = \exp\left( \frac{2\pi
 i p A}{N} \right)\, \Big( \C{P}^p \lst \C{C} \Ket \Big) \quad.
\end{equation}
For the cyclic group $\ZZ_N$ at hand, the projectors are defined by
\cite{Cornwell1}
\begin{subequations}
\label{eigen2}
\begin{equation}
\label{eigen2a}
 \C{P}^p = \frac{1}{N} \sum_{A=0}^{N-1} \Gamma^p\left( T^A \right)^*\,
 T^A = \frac{1}{N} \sum_{A=0}^{N-1} \exp\left( -\frac{2\pi i p A}{N}
 \right)\, T^A \quad,
\end{equation}
where the action of $T$ was defined in (\ref{cyclic12}). In
(\ref{eigen2a}), $p$ ranges between $0$ and $(N-1)$ and covers all
irreducible representations. The fact that $T$ is unitary implies that
the $\C{P}^p$ are Hermitean and have the projector property
\begin{equation}
\label{eigen2c}
 \C{P}^p \C{P}^q = \delta_{pq}\, \C{P}^p \quad.
\end{equation}
The sum over all $\C{P}^p$ gives the identity on the atomic state
space,
\begin{equation}
\label{eigen2d}
 \sum_{p=0}^{N-1} \C{P}^p = \Eins \quad.
\end{equation}
\end{subequations}
The completeness relation (\ref{eigen2d}) implies that the state space
of the atomic system may be spanned by a basis such that each of its
members transforms as a basis vector (\ref{eigen1}) of an irreducible
multiplet (In the present case, all multiplets are singlets, since all
$\Gamma^p$ are one-dimensional).

\begin{figure}[htb]
\begin{minipage}[b]{0.3\textwidth}
   \centering
   \includegraphics[width=1.\textwidth]{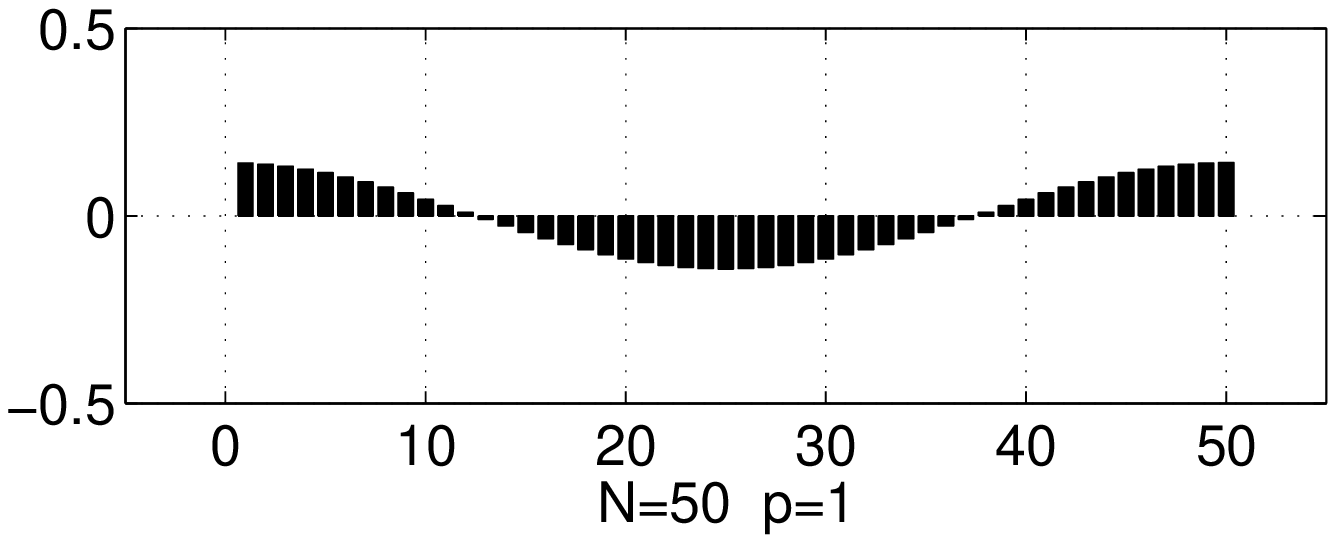}
\end{minipage} \\
\begin{minipage}[b]{0.3\textwidth}
   \centering
   \includegraphics[width=1.\textwidth]{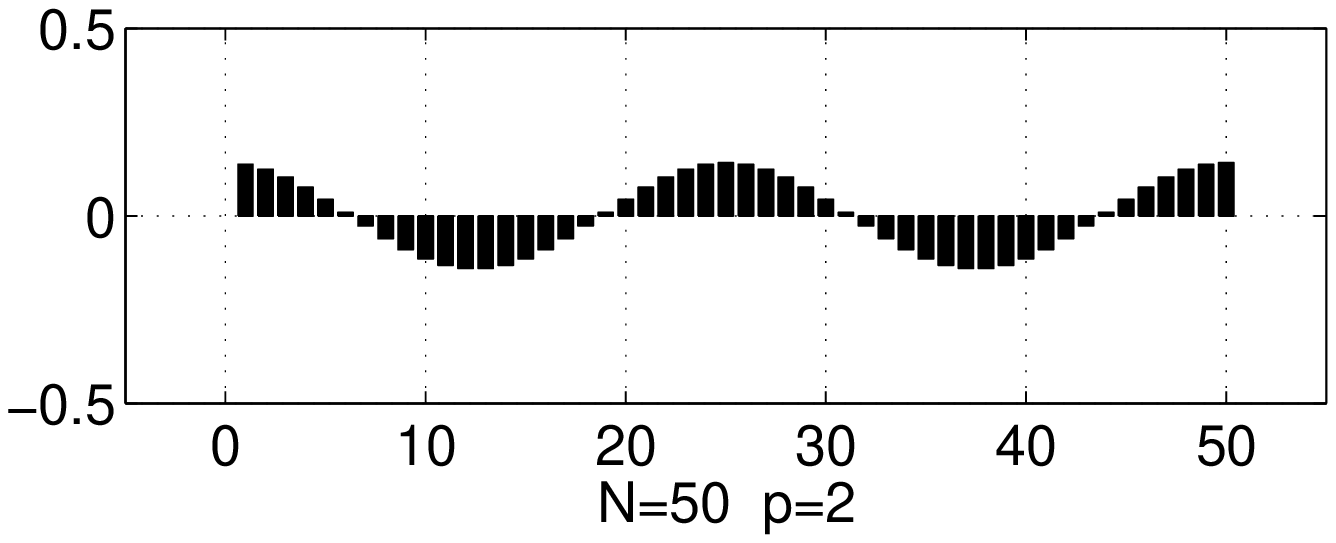}
\end{minipage} \\
\begin{minipage}[b]{0.3\textwidth}
   \centering
   \includegraphics[width=1.\textwidth]{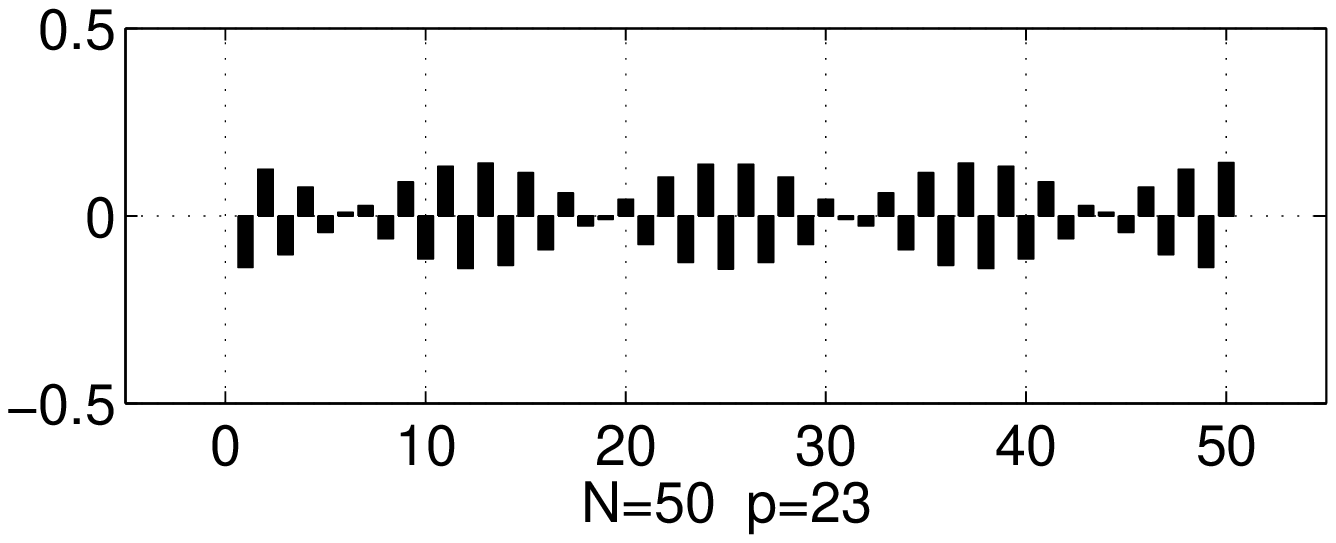}
\end{minipage} \\
\begin{minipage}[b]{0.3\textwidth}
   \centering
   \includegraphics[width=1.\textwidth]{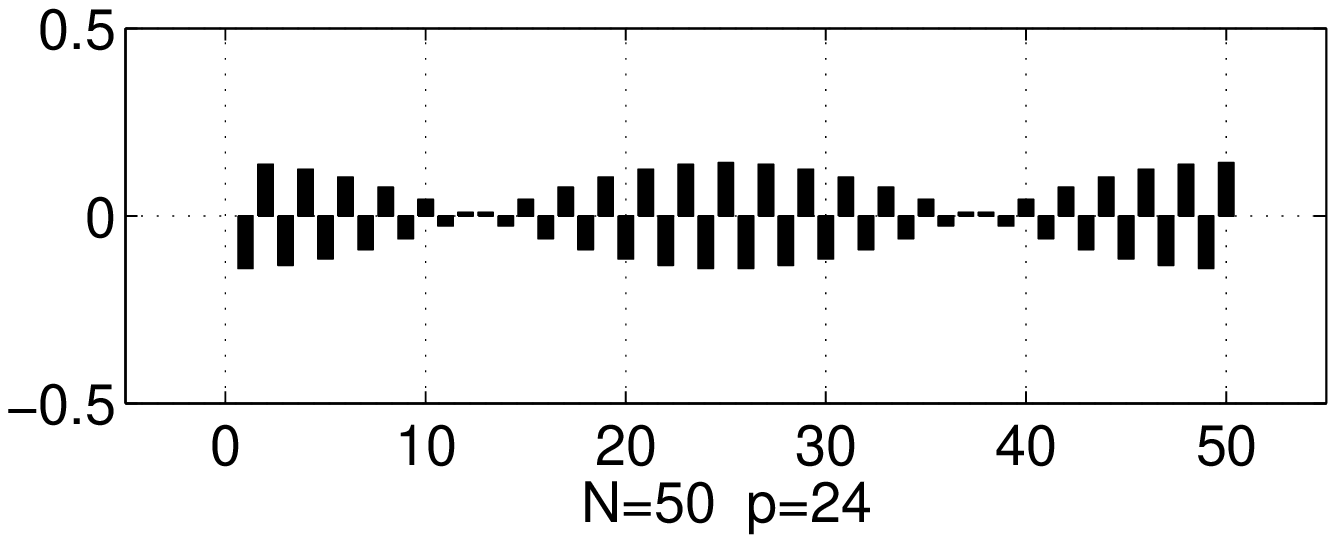}
\end{minipage} \\
\begin{minipage}[b]{0.3\textwidth}
   \centering
   \includegraphics[width=1.\textwidth]{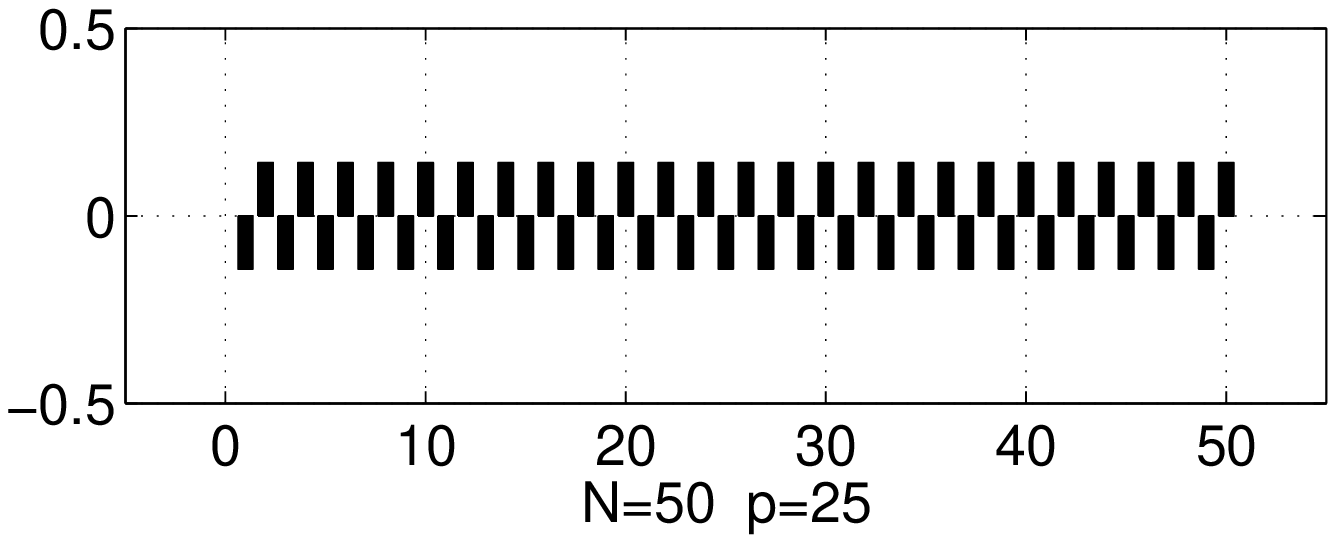}
\end{minipage}
\begin{minipage}[b]{0.3\textwidth}
   \centering
   \includegraphics[width=1.\textwidth]{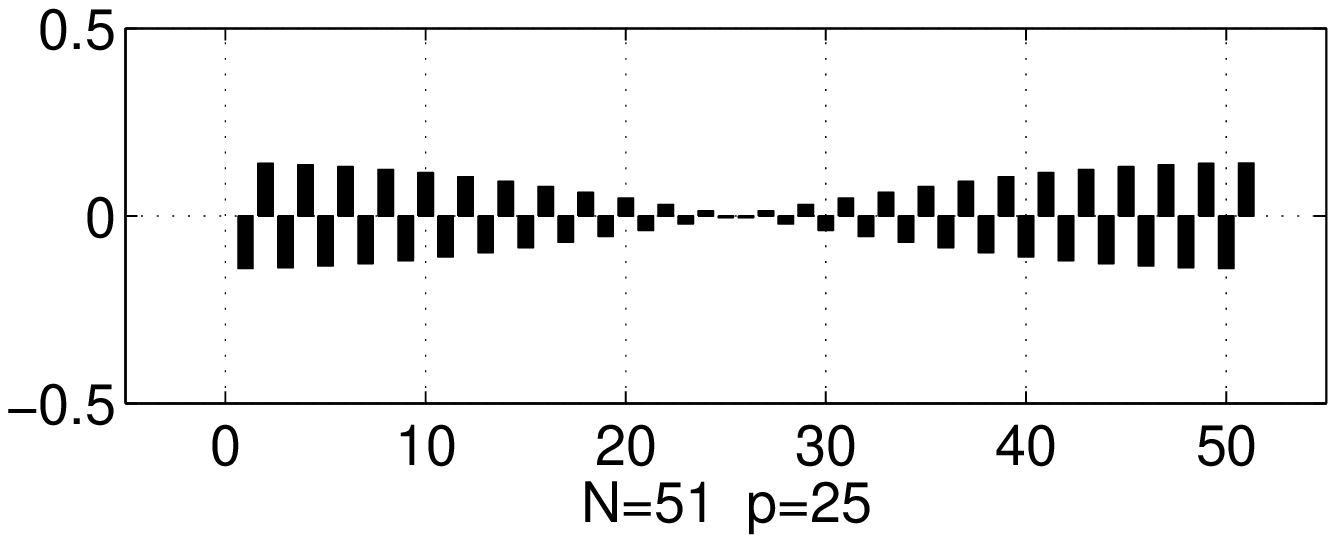}
\end{minipage}
\caption{The real parts $\Re c^p_A$ of the coefficients $c^p_A = \Bra
A,0 | \co{p} \Ket$ for the eigenstates $p=1,2,23,24,25$ of a
configuration with $N=50$ outer atoms, and for $p=25$ with $N=51$
outer atoms. Since $p \neq 0$, these coefficients are the same for
both configurations (a) and (b). \label{figure3}}
\end{figure}

We now construct such a basis in the single-excitation subspace of the
atomic state space which has dimension $(N+1)$ for configuration (a)
and $N$ for configuration (b). We first treat case (a): We start with
an arbitrary correlated state $\lst \co{ } \Ket$,
\begin{equation}
\label{eigen4}
 \lst \co{} \Ket = c_z |z,0\rangle + \sum_{A=1}^N c_A |A,0\rangle
 \quad,
\end{equation}
and apply the projector $\C{P}^p$; the result is
\begin{equation}
\label{eigen5}
 \C{P}^p\, \lst \co{} \Ket = \left\{ \frac{1}{N} \sum_{A=0}^{N-1} \exp
 \left( - \frac{2\pi i p A}{N} \right) \right\}\, c_z\, \lst z,0\Ket +
 \left\{ \frac{1}{N} \sum_{B=1}^N \exp\left( - \frac{2\pi i p B}{N}
 \right)\, c_B \right\}\; \times \sum_{A=1}^N \exp\left( \frac{2\pi i
 p A}{N} \right)\, \lst A,0 \Ket \quad.
\end{equation}
The factor before $c_z$ is equal to $\delta_{p0}$; thus we see that
the general form of a normalized basis vector carrying the irreducible
representation $\Gamma^p$ is
\begin{equation}
\label{eigen6}
 \C{P}^p \lst \co{} \Ket = \delta_{p0}\, c_z\, \lst z,0 \Ket + c
 \sum_{A=1}^N \exp \left( \frac{2\pi i p A}{N} \right)\, \lst A, 0\Ket
 \quad, \quad |c_z|^2 + |c|^2 = 1 \quad.
\end{equation}
It follows that, for $p \neq 0$, the eigenspaces of $T$ are
one-dimensional, and are spanned by basis vectors
\begin{equation}
\label{eigen7}
 \lst \co{p} \Ket \equiv \frac{1}{\sqrt{N}} \sum_{A=1}^N \exp \left(
 \frac{2\pi i p A}{N} \right)\, \lst A, 0\Ket \quad, \quad \bra \co{p}
 | \co{p} \ket = 1 \quad.
\end{equation}
This statement is true for both configurations (a) and (b). We
emphasize again that, here, $N$ denotes the number of {\it outer}
atoms. The coefficients $c^p_A = \Bra A,0 | \co{p} \Ket$ of states
(\ref{eigen7}) have the property that
\begin{equation}
\label{eigen7B}
 c_A^{N-p} = \left( c_A^p \right)^* \quad.
\end{equation}
In Fig.~\ref{figure3} we plot the real parts of $c^p_A$ for some of the
$p$-states with $N=50$ and $N=51$ outer atoms.

If $p=0$, the associated eigenspace $\msf{T}_0$ is one-dimensional for
configuration (b), and is spanned by
\begin{equation}
\label{eigen8}
 \lst \co{0} \Ket \equiv \frac{1}{\sqrt{N}} \sum_{A=1}^N \lst A, 0\Ket
 \quad.
\end{equation}
For configuration (a), the eigenspace $\msf{T}_{0}$ is
two-dimensional, with basis vectors $\lst \co{0} \Ket$ as in
(\ref{eigen8}) and a second basis vector
\begin{equation}
\label{eigen9}
 \lst \co{z} \Ket \equiv \lst z, 0\Ket \quad.
\end{equation}
On account of (\ref{eigen2c}), basis vectors pertaining to different
irreducible representations are automatically orthogonal. For the case
of the two-dimensional $p=0$ subspace of configuration (a) the two
basis elements may always be chosen as orthonormal, as we have done in
(\ref{eigen8}, \ref{eigen9}).

Thus we have accomplished the decomposition of the space of atomic
degrees of freedom into carrier spaces $\msf{T}_p$ for the irreducible
representations $\Gamma^p$ of the symmetry group $\ZZ_N$ for both
configurations (a) and (b). All that remains to be done now is to
diagonalize the channel Hamiltonian $\C{H}$ on each of these subspaces
separately.

\section{Diagonalization of the channel Hamiltonian on carrier spaces
 of the symmetry group}

\subsection{Configuration (b)}

We first discuss configuration (b) without central atom. In this case,
each of the eigenspaces of $T$ is one-dimensional, and is spanned by
states~(\ref{eigen7}, \ref{eigen8}). Since $\C{H}$ commutes with $T$,
it follows that each of the $\lst \co{p}\Ket$, $p=0, \ldots, N-1$, is
automatically a right eigenvector of $\C{H}$, or equivalently, of the
matrix $\C{R}$. The associated eigenvalue is found to be
\begin{equation}
\label{diago8}
 \mu_p = i + \sum_{A=2}^N M(k_{eg} R_{1A})\, \cos\left( {\frac{2\pi i
 p (A-1)}{N}} \right) \quad,
\end{equation}
which shows that
\begin{equation}
\label{diago12}
 \mu_{N-p} = \mu_p \quad,
\end{equation}
hence some of the eigenvalues are degenerate; the exact result is
presented in Table~\ref{TabDegeneracy}.
\begin{table}
\begin{center}
{ Level degeneracy for configurations (a) and (b)}
\end{center}
\begin{equation*}
\begin{array}{| @{\hspace{.5em}} r @{\hspace{.5em}} || @{\hspace{.5em}}
r@{}c@{}l @{\hspace{.1em}} @{\hspace{.5em}} | @{\hspace{.5em}} r@{}c@{}l
@{\hspace{.5em}} |} \hline
     & N & = & 2n+1 & N & = & 2n \\ \hline
 \hline
 \text{degenerate} & p & = & \begin{array}{r@{}c@{}l} 1, & \ldots, & n
 \\ 2n, & \ldots, & n+1 \end{array} & p & = & \begin{array}{r@{}c@{}l}
 1, & \ldots, & n-1 \\ 2n-1, & \ldots, & n+1 \end{array} \\ \hline
 \text{non-degenerate} & p & = & 0\, (\pm) & p & = & 0\, (\pm), n
 \\\hline
\end{array}
\end{equation*}
\caption{Degeneracy of the eigenvalues for quantum numbers $p=0 (\pm),
\ldots, N-1$, for odd and even $N$, for configurations (a) and
(b). Vertically stacked $p$ values in the second row are mutually
degenerate. \label{TabDegeneracy} }
\end{table}
On making use of~(\ref{eig7}) we finally find
\begin{equation}
\label{diago11}
 \mu_p = \sum_{A=2}^N S(k_{eg} R_{1A})\, \cos \left( \frac{2\pi p
 (A-1) }{N} \right) + i \Bigg\{ 1 + \sum_{A=2}^N D_1(k_{eg} R_{1A})\,
 \cos \left( \frac{2\pi p (A-1) }{N} \right) \Bigg\} \quad,
\end{equation}
where the real/imaginary parts contains the level shifts and decay
rates, respectively. A plot of these quantities for the first four
eigenvalues for $N=7$ outer atoms is given in Fig.~\ref{figure4},
where the approximated shift function $S_{\text{approx}}$ based on
eq.~(\ref{appr2}) has been used.

\subsection{Mechanisms for super- and subradiance}

\begin{figure}[htb]
\begin{minipage}[b]{0.45\textwidth}
   \centering \includegraphics[width=1.\textwidth]{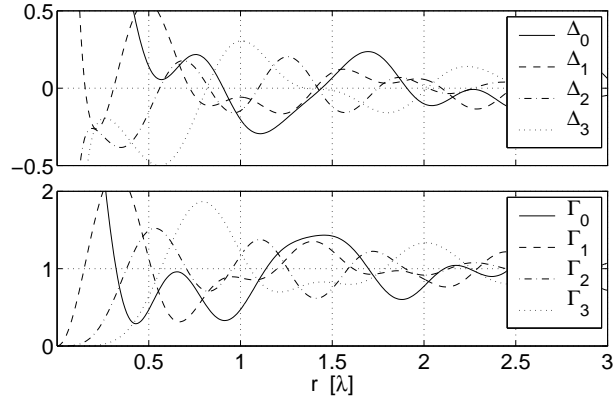}
\end{minipage}
\caption{Relative level shifts $\wt{\Delta}_p /\Gamma$ and decay rates
$\Gamma_p / \Gamma$ for the states $p=0,1,2,3$ with $N=7$ outer atoms
and no central atom. From the second plot we see that none of the
$p$-states can generally be declared as super- or subradiant; rather,
for each state there exist ranges of the radius $r$ for which the
state $\lst \co{p}\Ket$ is maximally super- or maximally
subradiant. The physical reason for this dependence lies in the
interference due to varying phase differences between spatially
retarded radiative contributions from all but one atom, at the site of
a given atom. \label{figure4}}
\end{figure}

The Figures~\ref{figure3} and~\ref{figure4} give some insight into
the mechanism of super- and subradiance. Studying first the limit of
vanishing radius $r \rightarrow 0$, hence vanishing interatomic
distance, we see in Fig.~\ref{figure4} that the $p=0$ state is
the only one which has a nonvanishing decay rate in this limit: All
atomic dipoles are aligned parallely in this state [see
eq.~(\ref{eigen8})], and vanishing distance on a length scale of
wavelength means that all radiation emitted coherently from the sample
must interfere completely constructively. Hence, the sample radiates
faster than a single atom by a factor $N$, since the decay rate is
\begin{equation}
\label{Decay0}
 \lim\limits_{r \rightarrow 0} \Gamma_{p=0} = N \cdot \Gamma \quad,
\end{equation}
as follows from eqs.~(\ref{diago11}) and~(\ref{eig1B1}). Conversely,
the suppression of spontaneous decay for the states with $p>0$ is a
consequence of the fact that the dipoles have alternating
orientations, see eq.~(\ref{eigen7}) and Fig.~\ref{figure3}, so that,
in the small-sample limit $r \rightarrow 0$, roughly one-half of the
atoms radiate in phase, while the other half has a phase difference of
$\pi$; hence
\begin{equation}
\label{DecayP}
 \lim\limits_{r \rightarrow 0} \Gamma_{p>0} = 0 \quad.
\end{equation}
This behaviour is exemplified by the plots of $\Gamma_1, \Gamma_2,
\Gamma_3$ in Fig.~\ref{figure4}. However, the destructive
interference in the $(p>0)$ states is independent of how adjacent
dipoles are oriented. For example, in the limit $r \ll \lambda$, the
decay rate should not be noticeably affected by rearranging the
dipoles in different patterns as long as the 50:50 ratio of
parallel-antiparallel dipoles is kept fixed. What {\it will} be
affected by such a redistribution is the level shift, since the
Coulomb interaction between the dipoles in the sample may change
towards more attraction or repulsion between the atoms. By this
mechanism we can explain the divergent behaviour of the level shifts
in the small-sample limit: The shift of the $p=0$ state always behaves
like
\begin{equation}
\label{Shift0}
 \lim\limits_{r \rightarrow 0} \Delta_{p=0} = + \infty \quad,
\end{equation}
which arises from the Coulomb repulsion of the parallely aligned
dipoles. On the other hand, the level shifts for the $(p>0)$ states
depend on the relative number of parallely aligned dipoles in the
immediate neighbourhood of a given dipole, or conversely, on the
degree of balancing the Coulomb repulsion by optimal pairing of
antiparallel dipoles. As a consequence, states for which $p$ is close
to zero always have positive level shifts, since the Coulomb repulsion
between adjacent dipoles is badly balanced, as seen in the first two
plots in Fig.~\ref{figure3}, and the behaviour of $\Delta_0, \Delta_1$
in Fig.~\ref{figure4}. On the other hand, states for which $p$ is
close to $N/2$ tend to have antiparallel orientation between adjacent
dipoles, hence the Coulomb interaction is now largely attractive,
which explains why these states have negative level shifts in the
limit $r \ll \lambda$. This is seen in the last four plots in
Fig.~\ref{figure3} and the behaviour of $\Delta_2, \Delta_3$ in
Fig.~\ref{figure4}.

The same mechanism clearly also governs the super- or subradiance of
the sample with finite interatomic distance. In this case the
information about the orientation of surrounding dipoles at sites
$\v{R}_A$ is contained in the transverse electric field, which arrives
at the site $\v{R}_1$ with a retardation $| \v{R}_A - \v{R}_1 |/
c$. Thus, in addition to the phase difference imparted by the
coefficients $c^p_A$, there is another contribution to the phase from
the spatial retardation, which accounts for the dependence of the
level shifts and decay rates on the radius $r$. Apart from this
additional complication, the physical mechanism determining whether a
given state is super- or subradiant is clearly the same as in the
small-sample limit, and can be traced back to the mutual interference
of the radiation emitted by each atom, arriving at a given site
$\v{R}_1$. This radiation is emitted coherently by the atoms on
account of the fact that the sample occupies a {\it pure} collective
state.

\subsection{Configuration (a)}

Now we turn to compute the complex energy eigenvalues and eigenvectors
for configuration (a) with central atom. For $p = 1, \ldots, N-1$, the
basis vectors carrying irreducible representations of the symmetry
group are the same as before, and are given in
eq.~(\ref{eigen7}). Consequently, they are also right eigenvectors of
the matrix $\C{R}$. The associated eigenvalues $\mu_p$ turn out to be
the same as for configuration (b) and thus are given by formulae
(\ref{diago8}, \ref{diago11}). This result means that the presence or
absence of the central atom makes no difference if the system occupies
one of the modes $\lst \co{p} \Ket$, $p = 1, \ldots, N-1$, since the
central atom is not occupied in the states~(\ref{eigen7}).

On the other hand, the eigenspace $\msf{T}_0$ of $T$ corresponding to
the $p=0$ representation is now two-dimensional and is spanned by
$\Gamma^0$-basis vectors (\ref{eigen8}) and (\ref{eigen9}). Since
$\msf{T}_0$ is preserved by $\C{H}(\wt{E}_i)$, we now need to
diagonalize the channel Hamiltonian in this subspace: The
representation of the matrix $\C{R}$ in the basis $\lst \co{0}\Ket,
\lst \co{z} \Ket$ is
\begin{equation}
\label{cona3}
 \C{R}' = \Big[ \C{R} \Big]_{\lst \co{0} \Ket, \lst \co{z} \Ket} = \left(
 \begin{array}{c|c} \sum_{A=1}^N \C{R}_{1A} & \sqrt{N}\, \C{R}_{1z}
 \\[8pt] \hline \rule{0pt}{15pt} \sqrt{N}\, \C{R}_{1z} & i
 \end{array} \right) \quad.
\end{equation}
Thus, we must diagonalize a non-Hermitean matrix of the form $\C{R}' =
\left( \begin{array}{cc} a & b \\ b & i \end{array} \right)$. The
diagonalization yields $\C{R}'\, \lst \co{0, \pm} \right) =
\mu_{0\pm}\, \lst \co{0, \pm} \right)$, where {\bf temporarily} we
have defined
\begin{equation}
\label{cona12}
 \mu_{0\pm} = \frac{a+i}{2} \pm \frac{1}{2} \sqrt{ (a-i)^2 + 4 b^2}
 \quad,
\end{equation}
and
\begin{subequations}
\label{cona7A}
\begin{gather}
 \lst \co{0+} \right) = \bigg(\, \frac{1}{\sqrt{1+ c^2}} \,, \;
 \frac{c}{ \sqrt{ 1+c^2}} \, \bigg)^T \quad, \quad \lst \co{0-}
 \right) = \bigg(\, - \frac{c}{\sqrt{1 +c^2}}\, , \; \frac{1}{ \sqrt{
 1+c^2}} \, \bigg)^T \quad, \label{cona7Aa} \\
 c = \frac{2 b}{a-i + \sqrt{ (a-i)^2 + 4 b^2}} \quad. \label{cona7Ac}
\end{gather}
\end{subequations}
These definitions will undergo some refinement, as we shall explain
below. The eigenvectors $\lst \co{0\pm} \Ket$ corresponding to the
column vectors~(\ref{cona7A}) then can be written in terms of a
complex "mixing angle" angle $\wh{\theta}$ such that
\begin{equation}
\label{cona7B}
 \lst \co{0+} \Ket = \cos\wh{\theta}\, \lst \co{0} \Ket + \sin
 \wh{\theta}\, \lst \co{z} \Ket \quad, \quad \lst \co{0-} \Ket = -
 \sin \wh{\theta}\, \lst \co{0} \Ket + \cos \wh{\theta}\, \lst \co{z}
 \Ket \quad,
\end{equation}
where $\tan \wh{\theta} = c$. Such an angle always exists, but is not
unique:
\begin{equation}
\label{cona9comment}
 \wh{\theta} = \frac{i}{2} \ln \frac{i+c}{i- c} \quad.
\end{equation}

The eigenvectors (\ref{cona7A}, \ref{cona7B}) are {\it not}
orthogonal, consistent with the fact that the matrix $\C{R}'$ is not
Hermitean. However, (\ref{cona7A}) form an orthonormal system together
with
\begin{equation}
\label{cona9B}
 \Bra \co{0+}^* \rst = \hspace{0.5ex} \cos\wh{\theta}\, \Bra \co{0}
 \rst + \sin\wh{\theta}\, \Bra \co{z} \rst \quad, \quad \Bra \co{0-}^*
 \rst = - \sin\wh{\theta}\, \Bra \co{0} \rst + \cos\wh{\theta}\, \Bra
 \co{z} \rst \quad.
\end{equation}
The eigenvectors (\ref{cona7B}) together with the duals (\ref{cona9B})
now satisfy generalized orthonormality relations according to
(\ref{chde11c}),
\begin{equation}
\label{cona9C}
 \left( \begin{array}{c} \Bra \co{0+}^* \rst \\ \Bra \co{0-}^* \rst
 \end{array} \right) \left( \begin{array}{cc} \lst \co{0+} \Ket & \lst
 \co{0-} \Ket \end{array} \right) = \left(
 \begin{array}{cc} 1 & 0 \\ 0 & 1 \end{array} \right) \quad,
\end{equation}
and completeness in the two-dimensional space $\msf{T}_0$ can be
expressed as
\begin{equation}
\label{cona9D}
 \lst \co{0+} \Ket \Bra \co{0+}^* \rst + \lst \co{0-} \Ket \Bra
 \co{0-}^* \rst = \Eins \big|_{\msf{T}_0} \quad.
\end{equation}

The eigenvalues $\mu_{0\pm}$ can be expressed in terms of the matrix
elements of $\C{R}'$,
\begin{equation}
\label{cona13}
 \mu_{0\pm} = i + \frac{1}{2} \sum_{A=2}^N M\left( k_{eg} R_{1A}
 \right) \pm \frac{1}{2} \sqrt{ \left[ \sum_{A=2}^N M \left( k_{eg}
 R_{1A} \right) \right]^2 + 4 N\, M^2\left( k_{eg} r \right) } \quad,
\end{equation}
where the function $M= S+i D_1$ was defined in eq.~(\ref{eig7}), and
$r$ is the radius of the circle.

The definition of eigenvalues $\mu_{0\pm}$ as given in
eqs.~(\ref{cona12}, \ref{cona13}) is not yet the final one,
however. In (\ref{cona12}) we have assigned $\mu_{0+}$ to the
"positive" square root of the complex number $(a-i)^2+4b^2$, i.e. the
square root with positive real part. From (\ref{eig6a}) we then see
that the $\mu_{0+}$ so defined always comes with a negative level
shift, and therefore the associated real energy $\wt{E}_i + \hbar
\wt{\Delta}_{0+}$ is always smaller than the energy associated with
$\mu_{0-}$. This state of affairs would be acceptable as long as the
levels would never cross; but crossing they do, as can be seen in
Fig.~\ref{figure5}:
\begin{figure}[htb]
\begin{minipage}[b]{0.47\textwidth}
   \centering \includegraphics[width=1.\textwidth]{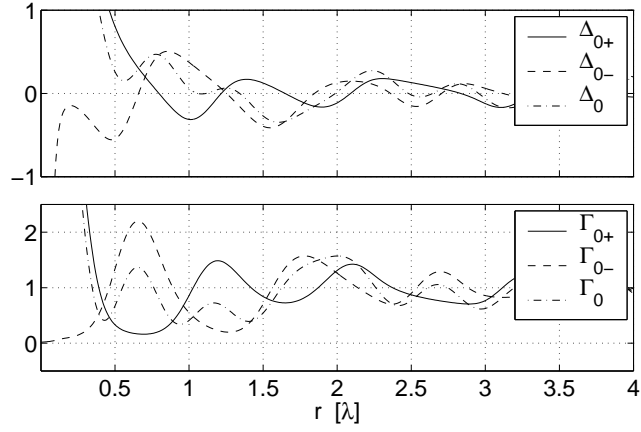}
\end{minipage}
\caption{For $N=10$ outer atoms, the level shifts $\wt{\Delta}_{0\pm}$
exhibit level crossing twice on every wavelength, for $r \ge
\lambda$. At the radius $r$ of a crossing, the beat
frequency~(\ref{rabi6}) vanishes, preventing any oscillatory
population transfer between $\lst \C{C}_0 \Ket$ and $\lst \C{C}_z
\Ket$. The states $\lst \C{C}_{0\pm} \Ket$ are defined such that, as
$r \rightarrow 0$, the level shifts $\wt{\Delta}_{0\pm}$ tend to $\pm
\infty$, respectively. --- Plotted are relative quantities
$\wt{\Delta}_{0\pm}/\Gamma$ and $\Gamma_{0\pm} / \Gamma$. For
comparison, level shifts and decay rates of the state $\lst \C{C}_0
\Ket$ without central atom are included. \label{figure5}}
\end{figure}
For radii greater than a certain lower bound $r_0$, which depends on
the number $N$ of outer atoms, we see two level-crossings per unit
wavelength. For $N=10$ this radius is roughly $r_0 \sim \lambda$. At
each crossing we have to reverse the assignment of square roots to
eigenvalues in order to obtain smooth eigenvalues. We therefore have
to redefine $\mu_{0\pm}$, and the associated eigenvectors, in order to
take account of this reversal at each crossing. The final task is then
to uniquely determine which of these smooth eigenvalues is to be
labelled $\mu_{0+}$ and $\mu_{0-}$. A look at Fig.~\ref{figure5} shows
that, for $r\le 0.7 \lambda$, no crossings occur, so that we can
uniquely identify the eigenvalues by their behaviour in the limit $r
\rightarrow 0$. In this spirit we finally define $\mu_{0\pm}$ to be
that eigenvalue whose associated level shift $\wt{\Delta}_{0\pm} = -
\Gamma \Re \mu_{0\pm}/2$ tends towards $\pm \infty$,
respectively. Physically, the dipoles in the associated state $\lst
\co{0+} \Ket / \left\| \co{0+} \right\|$ are aligned parallely, which
in the limit $r \rightarrow 0$ produces Coulomb repulsion, and hence
the positive level shift. It follows that the state $(p=0+)$ is the
natural analogue of the $(p=0)$ state in configuration (b), since they
both have the same behaviour at small radii. It is also expected to
decay much faster than a single atom, an expectation which is indeed
confirmed by the behaviour of the imaginary part $\Im \mu_{0+} =
\Gamma_{0+}/\Gamma$, which tends to $\sim N+1$ as $r \rightarrow
0$. Again this follows the pattern of the $p=0$ state in configuration
(b). A visual comparison between states $(p=0+)$ and $(p=0)$ is given in
Fig.~\ref{figure5} for $N=10$ outer atoms.

On the other hand, in the state $\lst \co{0-} \Ket / \left\| \co{0-}
\right\|$, the central atom is now oriented antiparallelly to the
common orientation of the outer dipoles, and, in the limit $r
\rightarrow 0$, is much stronger occupied than the outer atoms, as
follows from eq.~(\ref{cona7B}). Hence, as a consequence of Coulomb
attraction between the outer atoms and the central atom, the energy is
shifted towards $- \infty$, and at the same time, the system has
become extremely stable against spontaneous decay: This is reflected
in the fact that, as $r \rightarrow 0$, the decay rate $\Gamma_{0-}$
tends to zero as well.

As mentioned above, the states with higher quantum numbers $p=1,
\ldots, N-1$ are the same as in configuration (b), and have the same
eigenvalues. The two levels $p=(0\pm)$ are non-degenerate, except for
accidental degeneracy, and also are non-degenerate with the $(p>0)$
levels. As a consequence, the level degeneracy is similar to case (b),
and is again expressed in Table~\ref{TabDegeneracy}.

In a {\it Gedankenexperiment} we may think of switching off the
coupling of the central atom to the radiation field; in this case
$\C{R}_{Az}=0$, hence $\wh{\theta}= 0$, and the energy eigenstates
coincide with $\lst \co{0}\Ket$ and $\lst \co{z} \Ket$. As soon as the
central atom "feels" the radiation field, the true eigenstates are
(non-unitarily) rotated away from this basis. The modulus $|
\tan\wh{\theta} |$ of the tangens of the mixing angle $\wh{\theta}$
may be taken as a measure of the degree of correlation between the two
"unperturbed" states $\lst \co{0} \Ket$ and $\lst \co{z} \Ket$, or as
a measure of the strength of the interaction that couples the two
states. Alternatively, we could take the beat frequency $\omega_R$ of
the oscillation between the two unperturbed states as a correlation
measure:

\subsection{Quantum beats between unperturbed $p=0$ states}

As follows from eqs.~(\ref{cona7B}), the true eigenstates $\lst
\co{0\pm} \Ket$ are in general linear combinations of the
"unperturbed" irreducible basis vectors $\lst \co{0} \Ket$,
eq.~(\ref{eigen8}), and $\lst \co{z} \Ket$, eq.~(\ref{eigen9}); as a
consequence, the true eigenstates will give rise to quantum beats
between $\lst \co{0} \Ket$ and $\lst \co{z} \Ket$. In order to
determine the beat frequency we compute the amplitude for the
transition $\lst \co{z} \Ket \rightarrow \lst \co{0} \Ket$ [both
states being normalized], using eq.~(\ref{amp5AA1}),
\begin{equation}
\label{rabi1}
 \Bra \co{0} \right| U(t,0) \left| \co{z} \Ket = \Bra \co{0} | \co{0+}
 \Ket\, e^{- \frac{i}{\hbar} \Lambda_{0+} t }\, \Bra \co{0+}^* |
 \co{z} \Ket + \Bra \co{0} | \co{0-} \Ket\, e^{- \frac{i}{\hbar}
 \Lambda_{0-} t }\, \Bra \co{0-}^* | \co{z} \Ket \quad,
\end{equation}
where
\begin{equation}
\label{rabi2}
 \Lambda_{0\pm} = \wt{E}_i - \frac{\hbar \Gamma}{2}\, \mu_{0\pm}
 \quad,
\end{equation}
as follows from eqs.~(\ref{eig6}, \ref{amp6}). Using (\ref{cona7B},
\ref{cona9B}) and (\ref{rabi2}) we obtain
\begin{equation}
\label{rabi4}
 P_{\lst \co{z} \Ket \rightarrow \lst \co{0} \Ket}(t) = \Big| \Bra
 \co{0} \rst U(t,0) \lst \co{z} \Ket \Big|^2 = \left| \sin\wh{\theta}
 \cos\wh{\theta} \right|^2\, \Bigg\{ e^{- \Gamma_{0+} t} + e^{-
 \Gamma_{0-} t} - 2\, e^{-\frac{1}{2} \left( \Gamma_{0+} + \Gamma_{0-}
 \right) t }\, \cos \left[ \left( \wt{\Delta}_{0+} - \wt{\Delta}_{0-}
 \right) t \right] \Bigg\} \quad.
\end{equation}
The last term shows that the oscillation between the two states occurs
at the beat frequency
\begin{equation}
\label{rabi5}
 \omega_R  = \left| \wt{\Delta}_{0+} - \wt{\Delta}_{0-} \right| \quad.
\end{equation}
The beat frequency can be expressed in terms of the function $M$,
eq.~(\ref{eig7}), as
\begin{equation}
\label{rabi6}
 \omega_R = \frac{\Gamma}{2} \Re \sqrt{ \left[ \sum_{A=2}^N M \left(
 k_{eg} R_{1A} \right) \right]^2 + 4 N\, M^2\left( k_{eg} r \right) }
 \quad.
\end{equation}
\begin{figure}[htb]
\begin{minipage}[b]{0.4\textwidth}
   \centering \includegraphics[width=1.\textwidth]{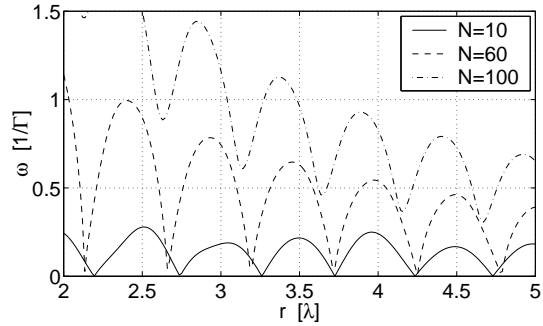}
\end{minipage}
\caption{The beat frequency eq.~(\ref{rabi6}) for three different
values $N=10,60,100$ of outer atoms, for radii between $2$ and $5$
wavelengths. At the radius of a level crossing $\wt{\Delta}_{0+} =
\wt{\Delta}_{0-}$, the beat frequency vanishes. It can be seen that
the greater the number of outer atoms $N$, the greater becomes the
radius $r_0$ beyond which level crossing occurs.
\label{figure6}}
\end{figure}

For a given number of outer atoms, there can exist discrete radii at
which the beat frequency $\omega_R$ vanishes. From
formula~(\ref{rabi5}) we see that this is the case precisely when the
two levels cross, and hence the normalized states $\lst \C{C}_{0+}
\Ket$ and $\lst \C{C}_{0,-} \Ket$ are degenerate in energy. This is
demonstrated in Figs.~\ref{figure5}~and~\ref{figure6}.

\begin{figure}[h]
\begin{minipage}[b]{0.4\textwidth}
   \centering \includegraphics[width=1.\textwidth]{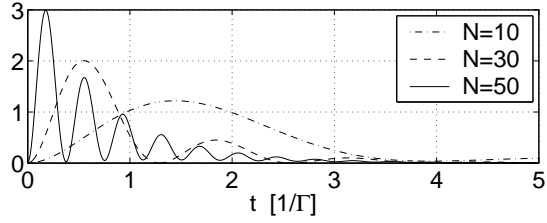}
\end{minipage}
\caption{The probability $P(t)/ | \sin \wh{\theta} \cos \wh{\theta}
  |^2$ for the transition $\lst \co{z} \Ket \rightarrow \lst \co{0}
  \Ket$, as given in eq.~(\ref{rabi4}), for $N=10,30,50$ outer
  atoms. The quantum beats for $N=10$ are practically invisible,
  making the transfer effectively aperiodic. \label{figure7}}
\end{figure}

The beat frequency $\omega_R$ so computed has to be treated with a
grain of salt, however. The reason is that the dynamics in the
radiationless $Q$-space does not preserve probability flux, since the
latter decays into the $P$-space when occupying the modes of outgoing
photons. This is reflected in the presence of damped exponentials in
formula~(\ref{rabi4}). Depending on the number of atoms involved and
the radius of the circle, this damping may be so strong, compared to
the amplitude of the beat oscillations, that the dynamical behaviour
effectively becomes {\it aperiodic}, i.e., exhibits no discernable
oscillations. An example is given by the $N=10$ plot in
Figure~\ref{figure7}.

\subsection{Analogy between $p=0$ states and hydrogen-like $s$ states}

It is interesting to note that for the $p\neq 0$ states, the central
atom is unoccupied, irrespective of the radius of the circle, or the
number of atoms in the configuration. This means that the central atom
takes part in the dynamics only in a $p=0$ state. This is strongly
reminiscent of the behaviour of single-particle wavefunctions in a
Coulomb potential, such as a spinless electron in a hydrogen atom: In
this case, the electronic wavefunction vanishes at the origin of the
coordinate system, i.e., at the center-of-symmetry of the potential,
for all states with orbital angular momentum quantum number greater
than zero. On the other hand, in the case of our planar atomic system,
the circular configurations also have a center-of-symmetry, namely the
center of the circle; and, as we have remarked earlier, we can
interpret the single-excitation correlated states $\lst \C{C}_p \Ket /
\left\| \C{C}_p \right\|$ as wavefunctions of a {\it single
quasi-particle} which is distributed over the set of discrete
locations $\v{R}_z, \v{R}_A$ corresponding to the sites of the
atoms. Then the amplitudes $\Bra A,0| \C{C}_p \Ket= c^p_A$ play a role
analogous to a spatial wavefunction $\Bra \v{x} |\psi\Ket =
\psi(\v{x})$; and just as the hydrogen-like wavefunctions vanish at
the origin for angular momentum quantum numbers $l \neq 0$
\cite{BransdenJoachainBuch2Ed}, so vanish our quasi-particle
wavefunctions at the central atom for all quantum numbers other than
$p=0$. In both cases, the associated wave functions are {\it
isotropic}: The $s$ states transform under the identity representation
($l=0$) of $SO(3)$ in the case of hydrogen, and states $\lst
\C{C}_{0(\pm)} \Ket$ under the identity representation $(p=0)$ of
$\ZZ_N$ in the case of our circular configurations. This means that
the quantum number $p$ is analogous to the angular momentum quantum
number $l$ in the central-potential problem; with hindsight, this is
not surprising, since both quantum numbers $p$ and $l$ are indices
which label the unitary irreducible representations of the associated
symmetry groups $\ZZ_N$ and $SO(3)$, and in both cases a {\it
rotational} symmetry is involved.

\section{Exponential photon trapping in the circular configuration}
\label{PhotonTrapping}

\begin{figure}[h]
\begin{minipage}[b]{1\textwidth}
\subfigure[Circular configuration, without central atom
   \label{figure8} ]{ \includegraphics[width=.45\textwidth]{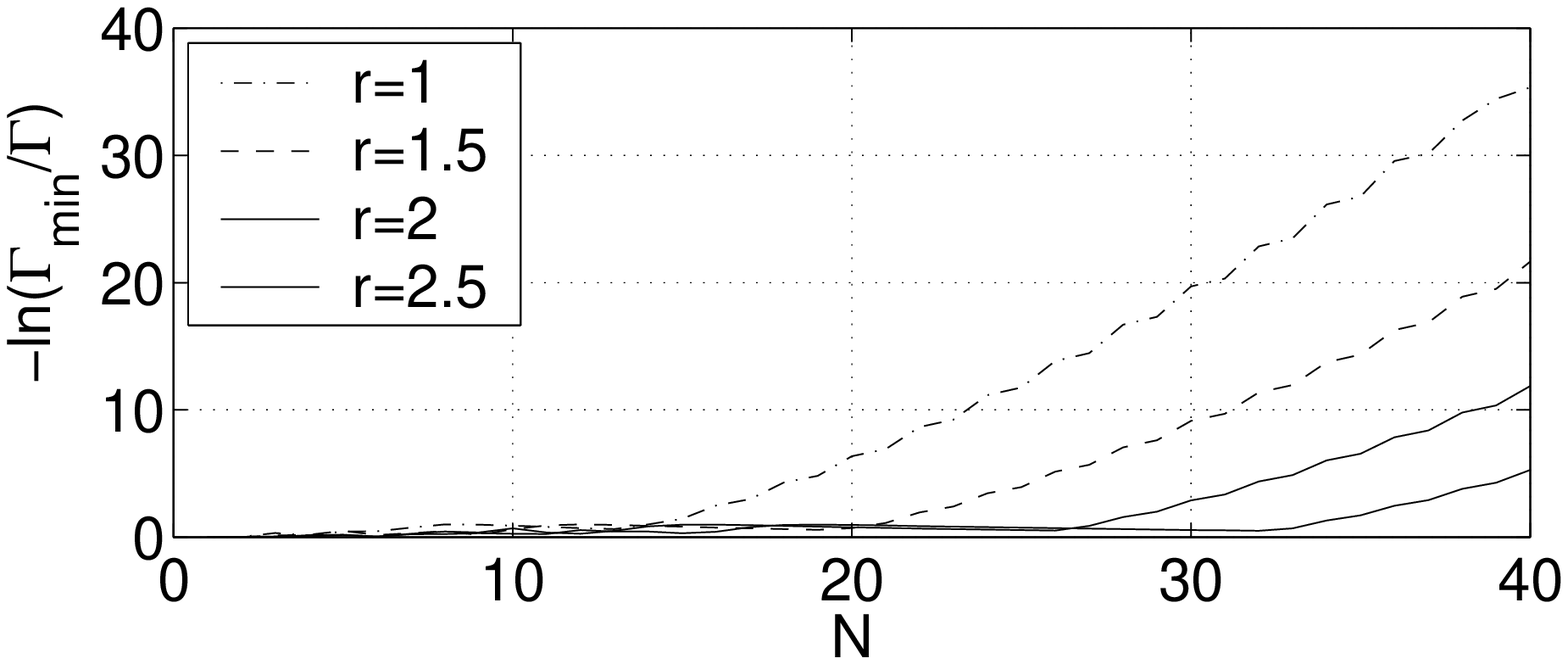}
   }
\subfigure[Linear-chain configuration) \label{figure10} ]{
   \includegraphics[width=.47\textwidth]{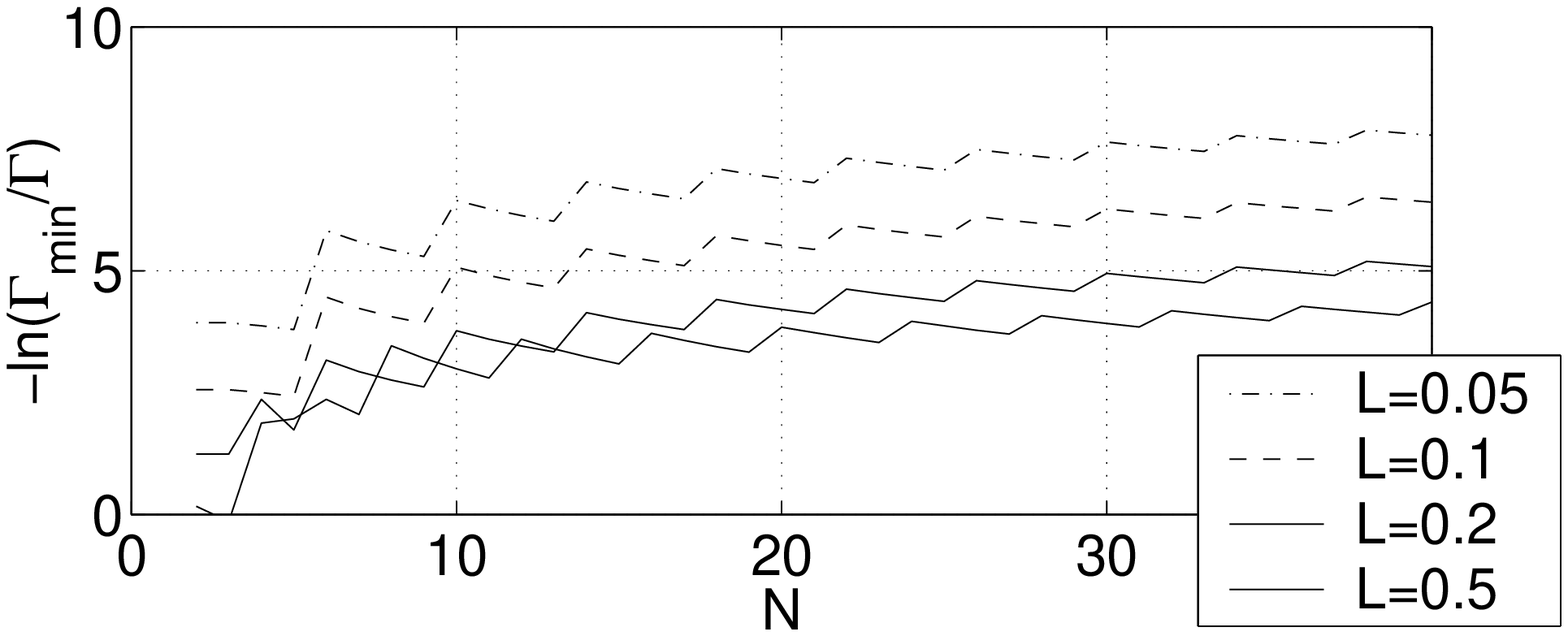} }
\end{minipage}
\caption{The negative logarithm of the minimal relative decay rate,
being proportional to the order of magnitude of suppression of
spontaneous decay, as a function of $N$, for fixed radii of the circle
in Fig.~\ref{figure8}, and for fixed length $L$ of the linear-chain,
in Fig.~\ref{figure10}. In Fig.~\ref{figure8} we see that for each
radius there exists a critical number $\hat{N}$ of atoms in the
configuration beyond which the order of magnitude by which spontaneous
emission is suppressed is roughly proportional to $N$. In the Figure,
these critical values are roughly at $\hat{N}=14, 20, 26, 33$
atoms. Beyond these values, the next-neighbour distance, which is
roughly equal to $2\pi r/N$, becomes smaller than the critical
distance $\lambda/2$. -- In Fig.~\ref{figure10}, the same analysis is
shown for four different lengths of the linear chain. A linear regime
is still visible, but both the slopes of the plots, as well as the
order of magnitude by which spontaneous decay is suppressed, is
visibly smaller than in the circular configuration.}
\end{figure}

In this section we are interested in the photon-trapping capability of
maximally subradiant states in the circular configuration, for large
numbers of atoms in the circle. To this end we choose a fixed radius,
increase the number of atoms in the configuration gradually, and, for
each number $N$, compute the decay rate $\Gamma_{\min}$ of the
maximally subradiant state for the given pair $(r,N)$, in
configuration (b) only. We then expect a more or less monotonic
decrease of $\Gamma_{\text{min}}$ as $N$ increases. But what precisely
is the law governing this decrease?  A numerical investigation gives
the following result: In Fig.~\ref{figure8} we plot the negative
logarithm $-\ln(\Gamma_{\text{min}} / \Gamma)$ of the minimal relative
decay rate at the radii $r=1, 1.5, 2, 2.5~\lambda$ for increasing
numbers of atoms. We see that from a certain number $N =\hat{N}$
onwards, which depends on the radius, $-\ln(\Gamma_{\text{min}} /
\Gamma)$ increases approximately linearly with $N$; in the figure, we
have roughly $\hat{N}=14$ for $r=\lambda$, $\hat{N}=20$ for
$r=1.5~\lambda$, $\hat{N}=26$ for $r=2~\lambda$, $\hat{N}=33$ for
$r=2.5~\lambda$. We also see that the slope is a function of the
radius $r$.

We note that the four curves in Fig.~\ref{figure8} imply the existence
of a common {\it critical interatomic distance}: For large $N$, the
next-neighbour distance $R_{nn}$ between two atoms on the perimeter of
the circle is roughly equal to $2\pi r /N$; if we compute this
distance for the pairs of values $(r,\hat{N})$ as found above we
obtain $R_{nn} = 0.45, 0.47, 0.48, 0.48~\lambda$, respectively. We see
that a critical distance of $R_c\approx 0.5~\lambda$ presents itself:
If, for fixed radius $r$, the number of atoms in the configuration is
increased, the next-neighbour distance $R_{nn}$ decreases; as soon as
$R=R_c$ is reached, the order of magnitude by which spontaneous
emission from the maximally subradiant state is suppressed becomes
approximately proportional to $N$.
Based upon this reasoning we see that, for a given radius $r$, the
critical number $\hat{N}$ of atoms in configuration~(b) is given by
$\hat{N} = \frac{4\pi r}{\lambda}$. Then,  we infer
from Fig.~\ref{figure8} that, approximately,
\begin{equation}
 \Gamma_{\text{min}} \simeq \Gamma \cdot e^{-s(r) (N- \hat{N})} \quad,
 \quad \text{for $N>\hat{N}$} \quad,
\label{ExpLaw1}
\end{equation}
where $s(r)$ determines the slopes of the curves in Fig.~\ref{figure8};
this function decreases monotonically with $r$. We must have the limit
$s(r) \xrightarrow{r\rightarrow\infty} 0$, because for large radii all
correlations between atoms must cease to exist, and hence
$\Gamma_{\text{min}} \rightarrow \Gamma$ in this limit. Let $\tau
\equiv 1/\Gamma$ denote the lifetime of the excited level in a single
two-level atom; then formula (\ref{ExpLaw1}) tells us that the
lifetime $\tau_{\text{min}}$ of a maximally subradiant state
pertaining to $\Gamma_{\text{min}}$ increases exponentially with the
number of atoms,
\begin{equation}
\label{ExpLaw2}
 \tau_{\min} = \tau\cdot e^{s(r) (N- \hat{N})} \quad, \quad \text{for
 $N>\hat{N}$} \quad.
\end{equation}

\section{Photon trapping in the linear chain configuration} \label{LinearChain}

\begin{figure}[t]
\begin{minipage}[b]{0.4\textwidth}
 \includegraphics[width=\textwidth]{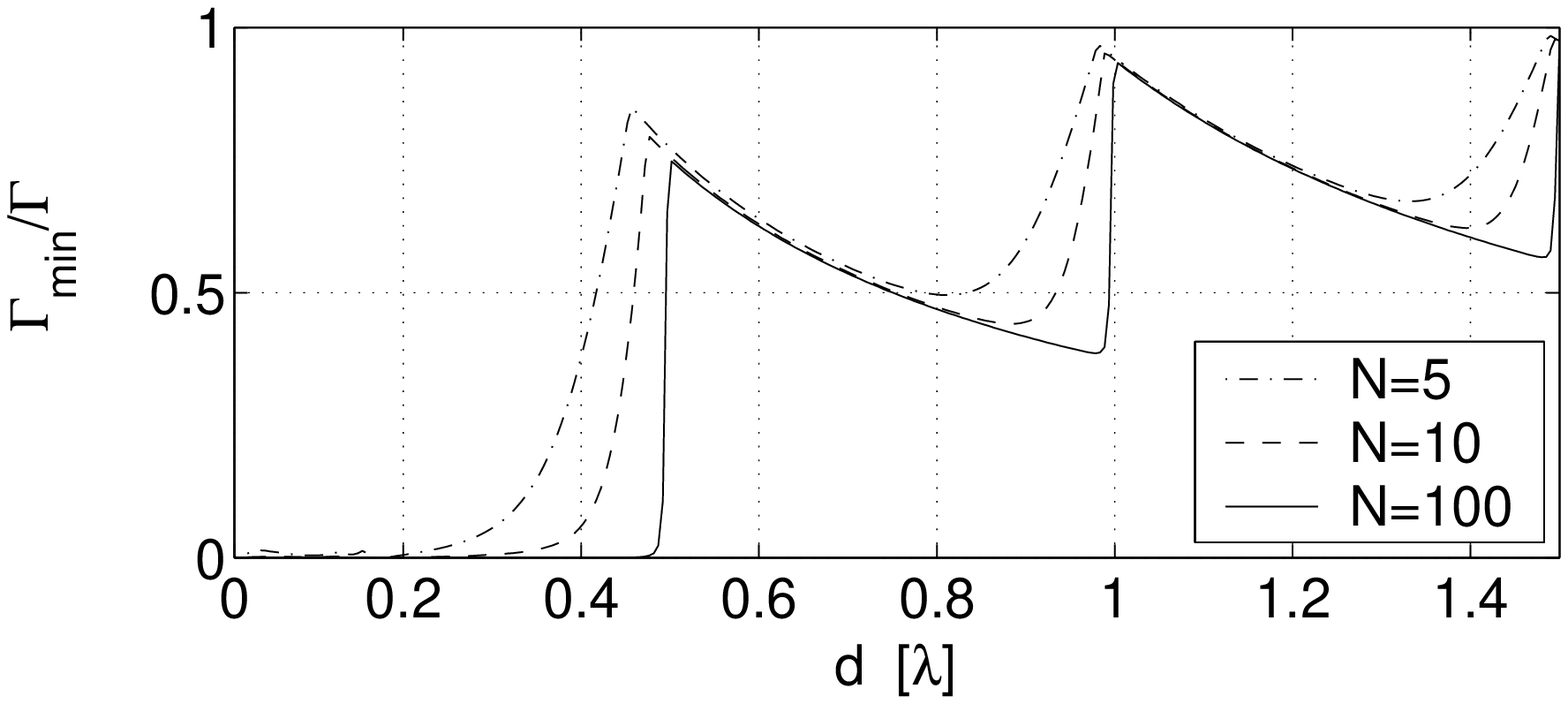}
\end{minipage} \hspace{1em}
\begin{minipage}[b]{0.4\textwidth}
 \includegraphics[width=\textwidth]{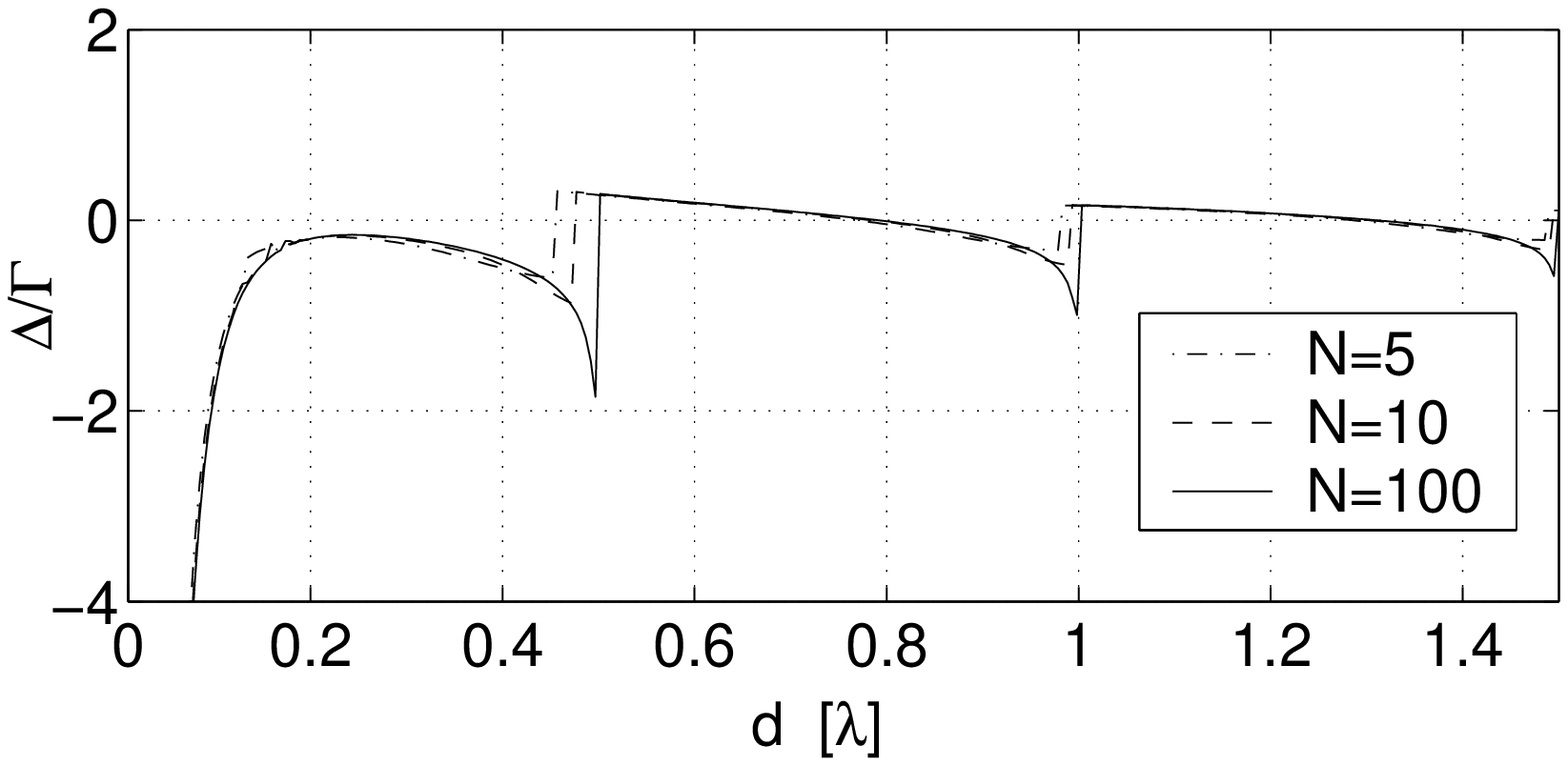}
\end{minipage}
\caption{The minimal decay rate in the linear chain configuration c),
for next-neighbour distances $d$ ranging between $0$ and $1.5\,
\lambda$, and $N=5,10,100$ atoms in the chain, respectively. For $d<
\lambda/2$ there is substantial suppression of spontaneous decay even
for relatively small numbers $N$ of atoms.
\label{figure9}}
\end{figure}

In this last section we study a configuration of $N$ identical
two-level systems which are arranged in a linear-chain
configuration. As before we focus attention on simply-excited states
only. This system no longer has a symmetry group, so that we have to
resort to numerical methods to compute energy eigenvectors and
eigenvalues. It turns out that the eigenvectors, i.e., the
coefficients of the correlated states $\lst \C{C}\Ket$ in the
uncorrelated basis $\lst A,0\Ket$, are qualitatively very similar to
those seen in Fig.~\ref{figure3} for the circular configurations,
except for some possible modifications at the boundaries of the
configuration. The energy eigenvalues exhibit a different behaviour
compared to the circular system, though. The most notable difference
is the occurrence of a region $d \in [0,d']$, $d' < \lambda/2$, in
which spontaneous decay can be strongly suppressed for the state
which, for the given distance $d$, has the minimal decay rate
$\Gamma_{\text{min}}$. This is shown in
Fig.~\ref{figure9}. Furthermore, in the limit $N \rightarrow \infty$,
the value of $d'$ approaches $\lambda/2$; and, as soon as
$d=\lambda/2$ is approached and exceeded, the minimal decay rate
exhibits a jump-like behaviour. This indicates that, even though
spontaneous decay is suppressed for $d< d'$, this suppression will
become increasingly unstable, and susceptible to environmental
perturbations, as soon as we approach the critical value
$\lambda/2$. In the region below this critical distance, spontaneous
decay may be suppressed by several orders of magnitude; for example,
for the three values of $N= 5, 10, 100$ as shown in
Fig.~\ref{figure9}, we have at the distance $d= 0.25\, \lambda$
minimal decay rates of $\Gamma_{\text{min}} \sim 10^{-2}, 10^{-3},
10^{-7}$, respectively.
However, the photon-trapping properties of the linear chain are not
nearly as pronounced as those of the circular configurations. In
Fig.~\ref{figure10} we plot the same quantities as in
Fig.~\ref{figure8} but for the linear chain with a prescribed length,
fixed at values of $L= 0.05, 0.1, 0.2, 0.5$ wavelengths. We see that a
linear regime exists here as well, but the suppression of spontaneous
decay is visibly smaller than in the circular configuration, as
$-\ln(\Gamma_{\text{min}}/\Gamma)$ does not exceed a value of $10$ in
the linear chain, while it reaches close to $40$ in the circle. To
further exemplify this point, let us prescribe a number of, say,
$N=40$ atoms, and demand a degree of suppression of spontaneous decay
of, say, $-\ln\left( \Gamma_{\text{min}}/\Gamma \right) = 5$. We then
ask: how small must the next-neighbour distance between atoms in the
circle on the one hand, and the linear-chain configuration on the
other hand, be in order to achieve the prescribed suppression of
radiative decay? A short computation shows that for the circle we need
a next-neighbour distance $R_{nn} \simeq 2\pi r/N \sim 0.4 \lambda$;
while in the linear-chain configuration, $d_{nn} \simeq 0.005 \lambda$
is required.  In other words, it is much harder to achieve the same
degree of photon trapping in the linear chain, and indeed, for the
values of parameters as given above, the circular configuration
performs better than the linear chain by a factor $0.4/0.005 = 80:1$!
Thus, if our objective is photon trapping, the circular configurations
definitely would be our first choice.

\section{Summary}

We develop the theory of simply-excited correlated states, their level
shifts and decay rates, for planar configurations of identical
two-level systems with parallel dipole moments, and apply the results
to circular and linear configurations of atoms. For the circular
systems, the atomic state space can be decomposed into carrier spaces
pertaining to the various irreducible representations of the symmetry
group $\ZZ_N$ of the system. Accordingly, the channel Hamiltonian on
the radiationless subspace can be diagonalized on each carrier space
separately, making an analytic computation of eigenvectors and
eigenvalues feasable. Each eigenvector can be uniquely labeled by the
index $p$ of this representation. For quantum numbers $p >0$ the
circular configuration is insensitive to the presence or absence of a
central atom, so that the wavefunction of the associated
quasi-particle describing the collective excitation of the sample
occupies the central atom only in a $p=0$ state. It is explained how
this feature is analogous to the behaviour of hydrogen-like $s$ states
in a central potential. The presence of a central atom in the circular
configurations causes level splitting and -crossing of the $p=0$
state, in which case damped quantum beats between two "extreme" $p=0$
configurations occur. For strong damping, the population transfer
between the two extreme configurations is effectively
aperiodic. Finally, a critical number of atoms, corresponding to a
next-neighbour distance of $\lambda/2$ on the circle, exists, beyond
which the lifetime of the maximally subradiant state increases
exponentially with the number of atoms in the circle. The significance
of this critical distance is exemplified by the jump-like behaviour of
the minimal decay rate in linear-chain configurations of atoms with
next-neighbour distance close to $\lambda/2$. It is demonstrated that
the photon-trapping capability of the circular system is pronouncedly
better than that of the linear-chain configurations.


\section{Appendix}

Here we prove a technical result which is used in the main part of the
paper:

Let $R_{1A}$ denote the distance between atoms $1$ and $A$ (i.e.,
outer atoms only). Let $F$ be any function of this distance, $F=
F(R_{1A})$. Let $p$ be an integer. Then
\begin{equation}
\label{append1}
 \sum_{A=2}^N F(R_{1A})\, \sin\left( \frac{2\pi p}{N} (A-1) \right) =
 0 \quad.
\end{equation}
{\it Proof:}
\nopagebreak

The sum $S$ on the left-hand side of (\ref{append1}) can be written as
\begin{equation}
\label{append2}
 S = \sum_{A=2}^N F(R_{1, N-A+2})\, \sin \left( \frac{2\pi p}{N}
 (N-A+1) \right) \quad.
\end{equation}
The sines are equal to
\begin{equation}
\label{append3}
 \sin \left( \frac{2\pi p}{N} (N-A+1) \right) = - \sin \left(
 \frac{2\pi p}{N} (A-1) \right) \quad,
\end{equation}
while the distances satisfy the equations
\begin{equation}
\label{append4}
 R_{1, N-A+2} = R_{1 A} \quad.
\end{equation}
If this is inserted into (\ref{append2}) we obtain an expression which
is the negative of (\ref{append1}), and as a consequence, $S$ must be
zero. \hfill $\blacksquare$


%
\acknowledgements{Hanno Hammer acknowledges support from EPSRC
grant~GR/86300/01.}



\end{document}